# Hierarchically Rough Surface used as Rewritable and Reprintable paper

Nitish Singh and Animangsu Ghatak*

A. Ghatak: Department of Chemical Engineering & Center for Environmental Science and Engineering, Indian Institute of Technology Kanpur, 208016

N. Singh: Department of Chemical Engineering, Indian Institute of Technology Kanpur, 208016 (India)

[*] Prof. A. Ghatak, Corresponding-Author, Author-Two, E-mail: aghatak@iitk.ac.in

**Abstract**

We describe here the design and fabrication of a rewritable and reprintable paper as an alternative to conventional papers which are generally used for writing and printing purposes. This paper is prepared by casting a crosslinkable silicone on a porous template, leading to hierarchical random patterns having dimensions ranging from tens of nano-meter to hundreds of micron. On the resultant rough surface, ink particles of wide ranging sizes and different surface chemical characteristics adhere via physical forces. Unlike conventional papers, in which particles get trapped inside the porous network of the cellulose fibers, here the ink particles remain accessible for cleaning via gentle shear with a rough and wet substrate. It is possible to write on this paper by different commercially available pens and to print using different printers. In almost all cases, the ink particles adhere to the paper surface strongly enough that they can't be dislodged by gentle shearing, yet adhesion can be decreased by inserting a liquid between the particles and the surface. Because of the resultant reversibility in adhesion, the paper remains usable over 200 cycles. This "reusable paper" is expected to be a viable alternative to the conventional papers that we use today.

**Introduction:**

Conventionally, paper consists of a porous mesh of cellulose fibers and some additives. During writing with a pen or printing by a printer, the colloidal particles or dye molecules, that constitute the ink, gets absorbed inside the three-dimensional mesh of the fibers. As a result, these particles no longer remain accessible to cleaning. So recycling of paper involves cutting, shredding and deinking by toxic chemicals and then re-pulping for making a new sheet of paper. Thus, making both fresh and recycled papers are expensive in terms of burden on the environment and consumption of energy, water, materials and time. The issue of environmental impact of using paper has been sought to be addressed in multiple ways, mostly involving ink material that either changes colour or gets vanished when exposed to an external field. These external agents include UV light[1-9], water[10], mechanical stress[11,12], heat[13,14] and pH[15]. Significant research has been done on all these methods to develop a viable and full-proof solution for large-scale manufacture and use of re-writable paper. For example, a supra-molecular hydrochromic dye system has been proposed[10] that undergoes red-shift when exposed to water. Besides that, different types of photochromatic dyes have been used to attain rewritability. For example, ink made of azobenzene metallic nanoparticles of gold and silver has been shown to undergo cis-trans isomerization triggered by UV-light and to revert back to its original form under daylight[1]. Polymer liquid crystals have been made to switch between out-of-plane and in-plane orientation when exposed to UV light[2]. In some cases, UV light has been used for writing whereas heating has been used for erasing the ink[3]. Photo-cyclization induced by UV light is another process that has allowed rewritability[8,9]. In most of these methods, however, legibility lasts for few hours and the writing-erasing cycle can be continued only for limited number of times. Stability of the organic dye molecule in presence of moisture and oxygen at the surroundings too is an issue. In several cases, the contrast of the writing is rather low in daylight condition[16-19]. Attempts have also been made to induce switching the colour of mechano-chromic molecules using mechanical force[11,12]. Similarly pH-sensitive materials, that change colour under the influence of acid-base reactions, too have been used for this purpose[15,20,21]. But, these methods are complex and many often they require handling of toxic acids and bases. Use of photonic crystals has also been proposed for rewriting purposes[22,24]; but it requires swelling of the substrate which is a slow process. Paper substrates based on swelling of the substrate by use of a hygroscopic salt[25] suffers from the possibility of evaporation of the solvent due to which the ink gets vanished over time[26]. Attempts have been made also to use photo-chromic response of electro-spun conjugated organic-inorganic hybrid membrane like that of PVP/a-$WO_3$ which changes color reversibly from being colorless in oxidizing environment to bluish in presence of external stimuli like electric field or UV light illumination. However, here the paper can only be used for UV writing and ozone erasing, which in essence demands use of specialized tools. To summarize, despite many advances, a viable ink that will allow large-scale use of rewritable paper doesn't exist; nor there exist any rewritable paper that will allow its adoption without requiring the habit of conventional writing to be changed altogether. Hardly any work exits that will allow "reprintability" of the paper via processes similar to conventional ones.

In the second approach, attempts have been made to develop process for removing the ink from the conventionally written and printed-papers without disintegrating its fibrous network[27-28]. However, these processes involve multiple steps, in many cases, these methods generate toxic by-products, and in most cases, the ink does not get fully removed, yet the fibrous network of the paper gets damaged. In another attempt, printed particles on paper were exposed to ultrafast lasers[29,30] of wavelength 532 and 1064 nm with 10 ps pulses. However, here the paper got damaged to the extent that it could not be reused. In another laser free process[31], the toner particles were first exposed to broad-spectrum of intense pulsed light (IPL) from a xenon lamp and then removed by gently wiping with ethanol. This process generated cracks on the paper. Furthermore, coloured toner particles couldn't be removed in this process. By this method, the paper could be used through at the most five cycles of printing and deprinting.

To address the above problem, we have developed a hierarchical, randomly rough surface of a low energy material like silicone. This surface allows writing by almost all conventional pens and printing with most conventional printers, yet allows also the ink to be removed when desired. Roughness pattern in combination with low energy of this surface causes the right extent of wettability and frictional characteristics, so that a line can be drawn on it using different types of pens, including one that uses a conducting ink. A line drawn on it, remains macroscopically continuous and prominent, yet the ink-pigment remains accessible for easy removal. Thus, the roughness patterns meet three different needs: enhance the wettability of the paper by most inks, enhance its ability to adhere to ink particles of wide ranging sizes, yet allow partial contact of adhesion, so that the adhered particles can be removed via gentle processes.

## RESULTS

**Fabrication and characterization of reusable paper substrate (RPS):** Figure 1(a,b) shows the process of making the RPS. A porous surface of hydrogel[32] (pHEMA) was used as a template for making the rough substrates. Thin films (30 μm - 1.36 mm) of silicone were cast against these templates to generate surfaces with different yet reproducible roughness (Figure S1, Supporting Information). A thin rod made of this material was subjected to stretching experiment, which yielded its Young's modulus, $E = 1.35\,\mathrm{N\ m^{-2}}$ (the Poisson ratio of most silicone materials can be taken as, $\nu = 0.5$). The surface of the RPS was characterized by surface energy and roughness. The former was estimated by the contact angle method in which a smooth featureless surface of the same silicone was first prepared by crosslinking a film with free surface. Following Owen and Wendt method[33], the surface energy of this silicone surface was estimated to be 24.93 mJ $m^{-2}$ (Figure S2, Supporting Information). On such a surface, the adhesion of particles (laser printer toner) with surface energy, $\gamma_{part} \approx 30\,\mathrm{mJ\ m^{-2}}$ (Figure S5, Supporting Information) was expected to yield an ensemble average of thermodynamic work of adhesion, defined as, $W = \gamma_{sur} + \gamma_{part} - \gamma_{sur-part} \approx 60\,\mathrm{mJ\ m^{-2}}$ (Figure S5, Supporting Information). The actual work of adhesion, $W_{act}$ was however expected to be different because of roughness of the surface.

Roughness of the RPS was characterized by scanning its surface using atomic force microscopy (AFM) and by optical profilometry (Figure 1c-e, Figure S3, Supporting Information). The data of height profile $h(x, y)$, thus obtained were used for obtaining the 2D power spectral density[34] (PSD) (Figure S4, Supporting Information) as shown in Figure 1(f). The PSD data for different scanned area over the whole range superimposed in the log-log plot, suggesting that the surface was self-affine with a single fractal dimension, $D_f$. The PSD data were fitted to a master line to yield the Hurst exponent ($\mathrm{slope} = -2H - 2$) and fractal dimension ($D_f = 3 - H$) as, $H = 0.89$ and $D_f = 2.11$ respectively. The PSD curve was used for obtaining the ensemble average of root mean square (RMS) roughness, $h_{\mathrm{rms}} = \langle h^2 \rangle^{1/2}$ of the surface (Supporting Information). For the surface as in figure 1(f), $h_{\mathrm{rms}}$ was estimated to be $1.37\,\mu\mathrm{m}$; by using different templates, $h_{\mathrm{rms}}$ could be varied over the range $0.39\,\mu\mathrm{m}$ to $8.0\,\mu\mathrm{m}$ (Table S1, S2, Supporting Information). The surface was characterized also by the maximum and minimum cut-off wave numbers: $q_{\max}$ and $q_{\min}$. In figure 1(f), $q_{\min} = 0.08$ μm$^{-1}$ and $q_{\max} = 418$ μm$^{-1}$ were associated with surface features of wavelength $\lambda \approx 80$ μm and 15 nm respectively. For a particle of diameter $d$, features with wavelength $\lambda >> d$, are expected to appear smooth; whereas, those with $\lambda \leq d$ are expected to affect the actual area of contact and the effective strength of adhesion. At these two limiting length-scales of roughness, the mechanism of adhesion too is expected to be different[35]. For a moderately soft layer, as the ones used here and for weak adhesive interaction like Van der Waals interaction, in the limit of wavelength, $\lambda \rightarrow d$, the adhesion stress within the area of contact exceeds the elastic stress developed in the adhesive. So the adhesion is essentially dictated by Johnson, Kendall and Roberts (JKR) mechanics[36] in which interaction within the area of contact is taken into consideration. In the other limit, $\lambda << d$, deformation of the layer becomes less prominent, so the adhesion stress outside the area of contact dominates, and the mode of adhesion is dictated by the Deryagin, Muller and Toporov (DMT) model[37]. The force necessary to remove a sphere of radius $R$ off a flat adherent in DMT model exceeds that as estimated by the JKR model: $F_{adh}\big|_{\mathrm{JKR}} = \frac{3}{4} F_{adh}\big|_{\mathrm{DMT}} = -\frac{3}{2}\pi R W$. Within these two limits, the effective adhesion, $W_{act}$ between the particle and the adherent is dictated by the distribution of amplitude of undulations of the surface as defined by its PSD and has been calculated differently by different authors[35,38-41].

**Particle adhesion on rough surfaces:** Adhesion of particles on randomly rough surfaces has been examined extensively in literature, most prominent among them are the models proposed by Persson et al[35,38] and Johnson et al[39-41]. Persson et al has obtained the probability distribution of contact stress for a rough surface with a given PSD, for a given value of thermodynamic work of adhesion, $W$ and elastic modulus, $E$. This model shows that $W_{act}$ varies non-monotonically with the dimensionless roughness parameter, $h_{\text{rms}}/\lambda$. With increase in $h_{\text{rms}}/\lambda$, $W_{act}/W$ first increases exceeding 1, attains a maxima for an intermediate $h_{\text{rms}}/\lambda$, following which it decreases to a very small value. Accordingly, the expected values of $W_{act}/W$ were estimated for adhesion of particles of different sizes (Supporting Information), on different reusable paper substrates (Figure 1g). The fractal dimension and the RMS roughness of these surfaces were varied: $D_f = 2.315-2.031$ and $h_{\text{rms}} = 2.14-0.7\,\mu$m respectively. For each surface, $W_{act}/W$ varied non-monotonically with the particle size: $W_{act}/W$ attained maxima for an intermediate range of particle size, $d = 0.1-1.0\,\mu$m. This enhancement was moderate for surface with $D_f \le 2.145$, however, was expected to increase significantly for $D_f$ exceeding $2.31$, as noted also by Persson et al[34-35]. A similar non-monotonic variation in adhesion of particles on surfaces with different random roughness has been predicted also by using the modified Johnson parameter model[39-41]. In this model, a single dimensionless parameter combines the effect of $W$, and the roughness characteristics of the surface: $H$, $h_{\text{rms}}$, $q_{\max}$ and $q_{\min}$ (Supporting Information, Figure S6). In other word, both these models suggest that a hierarchically rough surface is expected to adhere differently to particles of different size ranges. It is this aspect of adhesion on hierarchically rough surfaces that we have exploited in designing reusable paper substrates (RPS) suitable for writing/printing with inks of wide ranging characteristics.

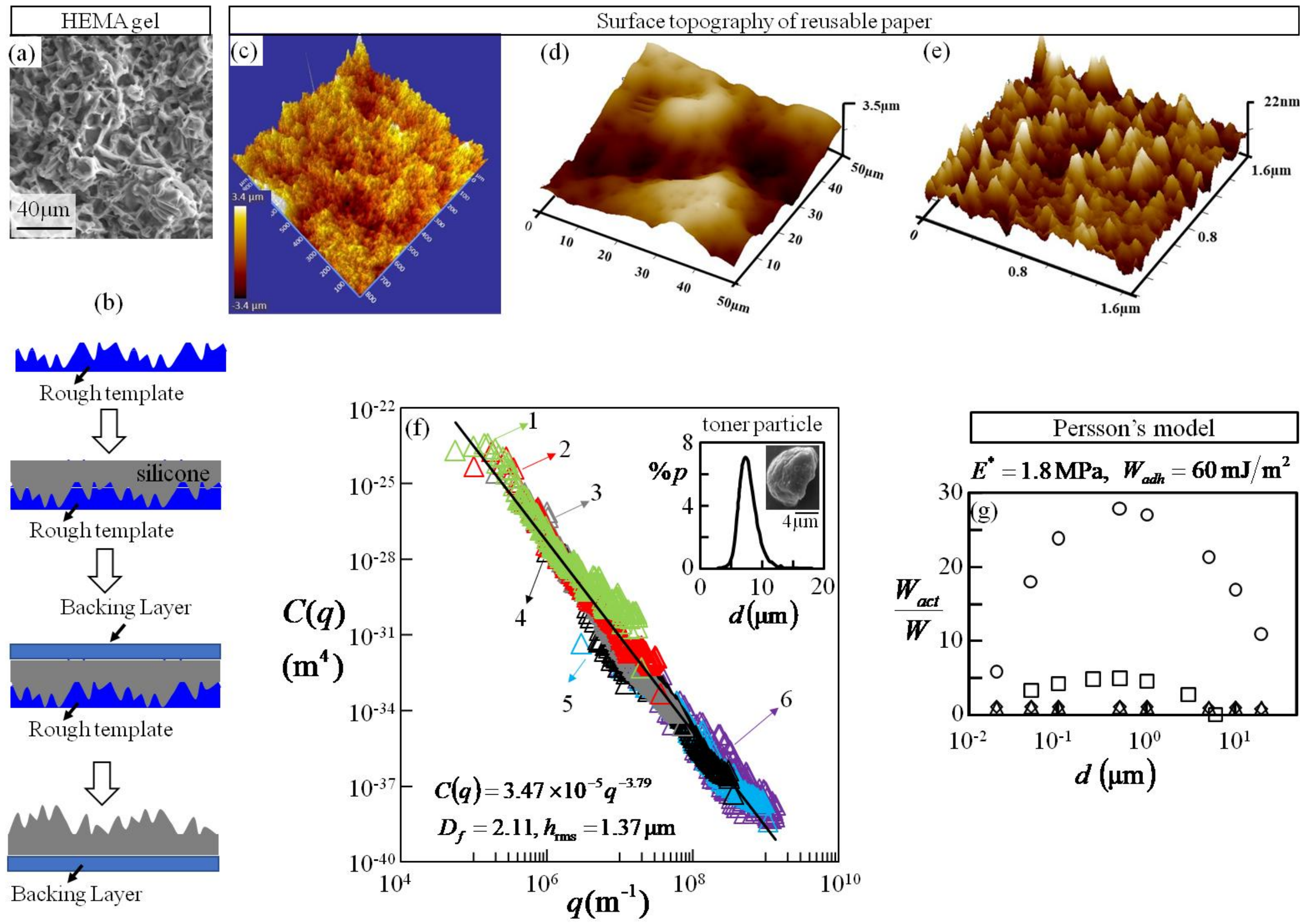

**Figure 1. The figure depicts preparation and characterization of the reusable paper substrate (RPS). (a) The Environmental Scanning Electron Microscope (ESEM) image represents the surface of a typical pHEMA gel prepared using monomer and water mixed at 10:90 w/w ratio. (b) The schematic depicts the process for making the RPS using the above template. The crosslinkable silicone liquid was poured over the template. A 75 GSM paper used as a backing was placed on the liquid silicone, which was crosslinked at room temperature for 6 hours. The crosslinked film on the backing was gently separated from the template. (c-e) The silicone surface was characterized by optical profilometry (scan area: 1.7 x 1.7 mm$^2$ to 0.17 x 0.17 μm$^2$) and atomic force microscopy (scan area: 90 x 90 μm$^2$ to 0.5 x 0.5 μm$^2$). (f) The plot shows two dimensional power spectral density (PSD) for**

**different scans all plotted in a single graph. The lines 1-5 correspond to scan sizes 90x90μm², 50x50μm², 25x25μm², 5x5μm², 1.7x1.7μm² and 0.5x0.5 μm² respectively. The solid line, representing the best fit of the data has a slope of** $-3.79$**. The inset shows the scanning electron micrograph of a typical laser toner particle and the size distribution of such particles. (g) The effective adhesion over thermodynamic work of adhesion,** $W_{act}/W$ **was calculated using Persson's model for different particle sizes for** $W = 60$ **mJ m**$^{-2}$ **and** $E/(1-\nu^2) = 1.8$ **MPa. Symbols ○, □, ◇ and △ represent respectively surfaces with** $D_f = 2.3$ **and** $h_{\text{rms}} = 1.8$ **μm, (**$2.145$**;** $1.3$ **μm), (**$2.11$**;** $1.08$ **μm) and (**$2.03$**;** $0.7$ **μm).**

**Discussions:**

**Adhesion of ink of conventional ball-point and sketch pens:** First, the RPS was subjected to writing and drawing with ball-point pens of different ink: black, blue, red and green (of three different makes). Dynamic light scattering experiments (Equipment used: Malvern Zetasizer Nano series Nano-ZS90, Malvern Instruments Ltd.) showed that the mean of the particle size distribution of the black and red ink were 1.5±0.5 μm and 0.2±0.1 μm respectively (Figure 2a, Figure S7, Supporting Information). These particles were large enough that they could not diffuse into the silicone network, but mostly remain on the surface; so, these ink particles were accessible for cleaning. The blue ink essentially consisted of dye molecules: phthalocyanine dye[42] and methyl blue[43] that could diffuse into the bulk of the silicone coating over time. In figure 2(a), we have plotted the expected $W_{act}/W$ values (represented by symbols) for adhesion of these particles on a surface with $D_f = 2.3$ (and $h_{\text{rms}} = 1.8$ μm), as estimated using Persson's model. The plots show that with decrease in the intrinsic work of adhesion $W$, $W_{act}/W$ decreases; consequently, the effective work of adhesion $W_{act}$ decreases too. $W$ can decrease when a layer of liquid gets inserted between the particles and the surface. Calculations show that even 10%w/w ethanol solution in water can decrease $W$ by 20 times to ~3 mJ/m$^2$ (Supporting Information). For a micron size particle of the black ink, the resultant $W_{act}$ can almost decrease to zero, thereby allowing its easy removal off the paper. The images of Figure 2(b) show that ball -point pens of different color and brands could be used sequentially on the RPS with intermediate erasing of the ink.

**Writing and erasing the ink of ball-point pen:** Images in Figure 2(c) (and in Figure S8, Supporting Information) depict erasing and writing with the black ball-point pen (Reynolds) over multiple cycles. In each cycle, a line was drawn on the RPS and was allowed to age for @6 hours. At this, the ink got dry enough that the ink particles couldn't be dislodged by gentle rubbing with finger (Figure S9, Movie S3, Supporting Information). However, when a cotton cloth wetted with water was rubbed over it, the particles of black and red ink got removed completely (Movies S1 and S2, Supporting Information). For blue ink, a fine trace remained stuck to the surface because of possible diffusion of the dye into the silicone network. The wet surface of the RPS was dried and was used again for writing. This sequence of writing and erasing could be continued over 100 cycles. To examine if writing resulted in any change in the surface roughness, the RPS was scanned under optical profilometer after every cycle; even after 100 cycles, the roughness and fractal dimension of the surface remained unaltered at $h_{\mathrm{rms}} = 1.34\,\mu\mathrm{m}$ (Figure 2d,e) and $D_f = 2.11$ respectively.

**Effect of surface roughness on writing with ball-point pen:** To examine the effect of roughness of the surface (varied from $h_{\mathrm{rms}} = 0.017$ μm to 12 μm) on their writability, in figure 2(f) (and in Figure S10, Supporting Information) we present the magnified image of a line drawn using a black ball-point pen (Goldex Klassy) on different surfaces. We have presented also the image of a line drawn on a conventional (75 GSM) paper. These images show that for $h_{\mathrm{rms}} < 0.4$ μm, the surface does not impart any pinning effect, so the thin film of the ink undergoes Plateau Rayleigh instability[44] and breaks up into tiny droplets. For surfaces with $h_{\mathrm{rms}} > 2$ μm, the surface consists of large craters which lead to pinning of ink and breaking of the line into islands of liquid. Pinning of the three phase contact line of a liquid front spreading on a heterogeneous surface has been examined by Joanny and De Gennes, by equating the effect of hysteretic force of the surface to the capillary force of the liquid[45]. Following their method (Supporting Information), the threshold value of surface slope $h'\big|_c$ , for pinning to occur for the ink of a ball-point pen, is estimated to be ~0.21. The RMS slope $h'_{\mathrm{rms}}$ calculated from the PSD of different surfaces (Figure S2, Supporting Information) show that for $h_{\mathrm{rms}} > 2$ μm, $h'_{rms}$ far exceeds this threshold limit. In contrast, for surfaces with $0.4 < h_{\mathrm{rms}}(\mu\mathrm{m}) < 2$, $h'_{rms} \sim 0.21$ and a continuous line could be drawn with high fidelity.

**Fidelity of writing with ball-point pen and the effect of ageing:** Fidelity of writing was quantified by considering two different parameters: width, $w$ of the line drawn on it and the surface coverage, $\%A$ of the ink within the area of this line (Figure 2g). In Figure 2(g), $w$ data normalized with $w_0$ for a conventional 75GSM paper, was presented against $h_{\mathrm{rms}}$ of RPS surfaces. Large value of $w/w_0 = 1.5 - 2.0$ signified that the ink could spread well on the RPS during writing, yet due to hysteresis, it did not retract back. $\%A$ was also estimated for different surfaces. For smooth surface with $h_{\mathrm{rms}} = 0.017$ μm, $\%A$ was ~40%, but increased to ~95% for RPS with intermediate roughness: $1.6 < h_{\mathrm{rms}}(\mu\mathrm{m}) < 2.2$. Hence, these surfaces were most suitable for writing with the ball-point pens. A similar set of experiments showed negligible ageing for writing with a black ball-point pen on a given RPS (Figure S12, Supporting Information).

**Writing and drawing with sketch pen:** Similar to ball-point pens, sketch pens of different colors: black, blue, green and orange could also be used for both writing and drawing reversibly over multiple cycles on the reusable paper substrates (RPS) (Figures S13-S15, Movie S10, Supporting Information).

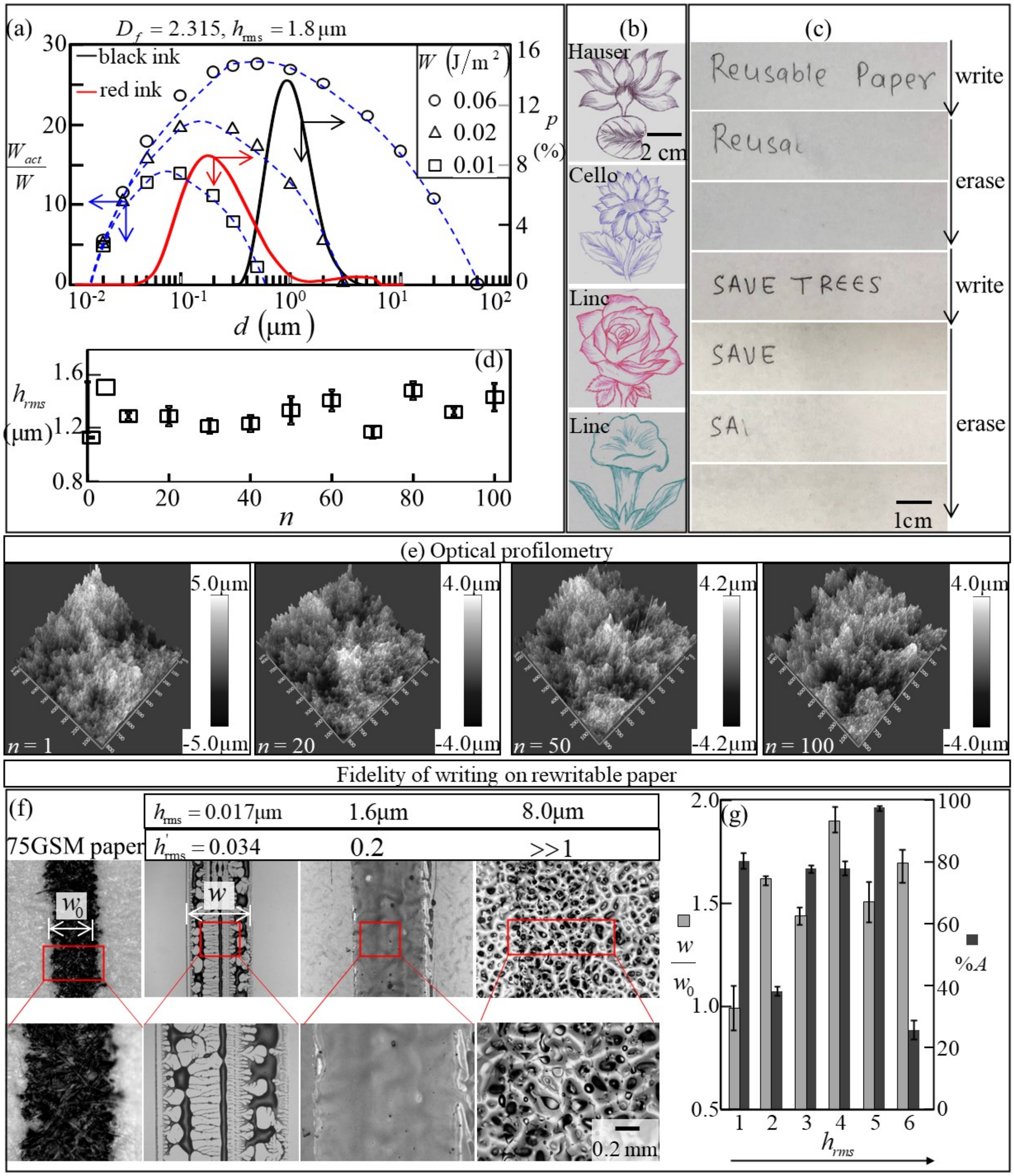

**Figure 2. The figure demonstrates reusability of an RPS with ball-point pen. (a) The symbols in the plot represent enhancement of adhesion: $W_{act}/W$ of particles of different sizes on a typical RPS ($D_f = 2.31$, $h_{rms} = 1.8\,\mu m$ and $E/(1-\nu^2) = 1.8$ MPa). $W_{act}/W$ was**

**calculated for $W = 0.06 - 0.01$ mJ/m² using Persson's model. In this plot, the continuous lines represent the particle size distribution of ink of red and black ball-point pen. (b) Ball-point pens of different color and make: Black (Hauser), blue (Cello), Red (Linc) and Green (Linc) were used sequentially for drawing on a rewritable paper ($h_{rms} = 1.6$ μm). (c) The sequence of images shows writing and erasing over repeated cycles. In each cycle, a black Reynolds ball-point pen was used for writing on the RPS, the ink was first dried in atmospheric condition and then wiped off using a wet cloth; the wet surface was dried again (inside a hot air oven at 50°C for 10 min) to make it suitable for writing. (d, e) The figures show optical profilometry images of the surface over 100 such cycles and the corresponding $h_{rms}$. (f, g) Fidelity of writing is demonstrated by drawing a line (using a black Goldex Klassy ball-point pen) on papers of different roughness, followed by capturing its magnified image under an optical microscope. The bar chart shows the width, $w$ of lines (normalized by that for a conventional paper): $w/w_0$ and the percentage area $\%A$ of ink coverage for different paper substrates. Bar 1 represents a conventional paper ($h_{rms}$ = 5.4μm); bars 2-7 correspond to RPS with $h_{rms}$ = 0.017, 0.39, 1.6, 2.0, 2.2 and 8.0 μm respectively.**

**Adhesion of toner particles of laser jet printer on reusable paper:** The reusability of RPS was examined by printing using a conventional laser jet printer. The black toner (Figure S16, Supporting Information) of this printer consisted of particles of size: 7±1 μm and aspect ratio: 1.5±0.2 (inset of Figure 1f). In laser printing, adhesion of particles is enhanced by creating static charges on the paper surface. This process leads to electrostatic interaction and irreversible bonding of the toner particles with the mesh of cellulose fibers of the paper[46,47]. However, the surface of the RPS is chemically and topographically different from that of a conventional paper. Therefore, the effectiveness and stability of the print on it was examined by printing lines and images and then finding the strength of adhesion of the toner particles with the RPS surface. Among many different toners, particles of hp 88a black laser jet toner cartridge (used in HP Laser Jet Pro MFP M126nw printer) bonded well with the RPS, such that gentle rubbing could not dislodge the particles from the surface. Using this printer, the RPS was first printed over a desired area; a scotch tape was placed over this printed surface and was brought in complete contact with the particle layer by a gentle press. The RPS was bonded to a rigid plate attached to a load cell and the tape was peeled at an angle of 90° at different peeling rate: 0.006 - 0.6 mm $sec^{-1}$ using a motorized drive (Figure 3a). For this peel angle, the debonding energy, $G$ equates to the peel force $F$ per unit width of the tape. At peeling rate <0.015 mm $sec^{-1}$, the toner particles didn't detach at all from the surface and the tape gently debonded from the top surface of the layer of particles. However, with increased rate of peeling, the particles detached to an increasing extent; complete detachment happened at peeling rate >0.3 mm $sec^{-1}$ (Movie S4, Supporting Information). The plots in figure 3(b) show the variation of fractional area $f$ through which the particles detached from the surface and that of the debonding energy $G$ with the peeling rate, $v$; both increased with $v$ and attained an asymptotic plateau value. In each case, the experiments were repeated over 8-10 samples, to obtain the representative data. The adhesive used in a typical scotch tape is known to have strain rate dependent rheology. Because of this complex material behavior, the mode of its debonding, from an adherent, alters with rate of peeling. At lower peeling rate, the adhesive behaves more like a viscous liquid, for which, dissipation occurs at its bulk. At higher rate of peeling, it behaves like an elastic solid, for which, energy remains conserved and so dissipation can happen at the interface with an adherent; thus more energy gets released at the interface, which drives the particles to debond from the paper. The debonding energy, $G$ as high as 1.0 J m$^{-2}$, can be attributed to electrostatic interaction between the toner particles and the surface. $G$ was high enough that the print was not susceptible to gentle shear, yet was low enough that the ink could be completely removed when sufficient pull-off load was applied (Movie S5, Supporting Information).

**Method of erasing ink of laser jet printer:** The toner particles of laser jet printer could be dislodged from the RPS by three different methods. For RPS with low roughness, $h_{\mathrm{rms}} < 1.6\,\mu\mathrm{m}$ and relatively smaller printed area: 1.8 cm x 5 cm, peel-off using a scotch tape was an effective way of removing the particles adhered to the RPS. However, this method was unsuitable when print over a large area, as that of an A4 size paper, was to be cleaned. Gentle wiping with a damp cloth soaked in ethanol was a way of removing these particles for RPS with wide ranging roughness and surface area (Movie S6, Supporting Information). The ink could be removed also by using a spin scrubber along with a continuous layer of ethanol over the paper surface (Movie S7, Supporting Information). By these methods, the surface could be completely cleaned off the ink as confirmed by SEM of the surface before and after cleaning off the ink (Figure S17 - S19, Supporting Information).

**Reprintability of the RPS with laser jet printer:** Figure 3(c) depicts printability of the paper over 50 cycles. To have a quantitative estimate of the print quality, the gray scale intensity along a line drawn across a printed image was obtained and the dimensionless resolution $(I_0 - I)/I_0$ ($I_0$, $I$: intensity at printed and un-printed portions respectively) was calculated. Irrespective of the number of cycles, the resolution and the fractional area covered by toner particles were almost constant (Figure S15(b), supporting information). The resolution of print on RPS was also same as that for conventional paper.

**Effect of ageing and surface roughness on adhesion of laser toner particles on RPS:** Figure 3(e) depicts the effect of ageing of laser toner particles printed on the RPS. A line was printed on the RPS ($h_{\text{rms}} = 1.6\,\mu\text{m}$, $D_f \sim 2.11$), the printed ink was removed by gentle peeling of a scotch tape after different ageing time:. The printed line and the portion of the paper from where the ink was removed, both were examined under optical microscope. The images in figure 3(e) show that the print quality and the effectiveness of removal of the ink, both remain unaltered over 30 days, signifying long-term stability of the print on RPS. Finally, the effect of roughness was examined by laser printing on RPS of different roughness: $h_{\text{rms}} = 0.017\,\mu\text{m}$ to 8.7 μm (figure 3(f)). The print was then removed by peeling off a tape at 0.15 mm/sec, similar to the experiment of figure 3(a). The resultant force vs. displacement ($F$ vs. $\Delta$) data consisted of a peak corresponding to initiation of the peeling front; following initiation, the force during propagation was nearly constant. For both $h_{\text{rms}} = 0.017\,\mu\text{m}$ and $1.6\,\mu\text{m}$, the propagation force was nearly identical, although, at this rate of peeling, the ink particles could be completely removed only for surface with $h_{\text{rms}} = 0.017\,\mu\text{m}$. For higher roughness, e.g. $h_{\text{rms}} = 8.7\,\mu\text{m}$, the adhesive on scotch tape could not access the particles adhered to inside the groove of the substrate, so the particles were hardly removed. It is worth noting also that RPS with smaller roughness, e.g. $h_{\text{rms}} = 0.017\,\mu\text{m}$, was incompatible with the paper driving mechanism of conventional printers. In essence, RPS with intermediate roughness: $h_{\text{rms}} = 1.6\,\mu\text{m}$, was most suitable for printing.

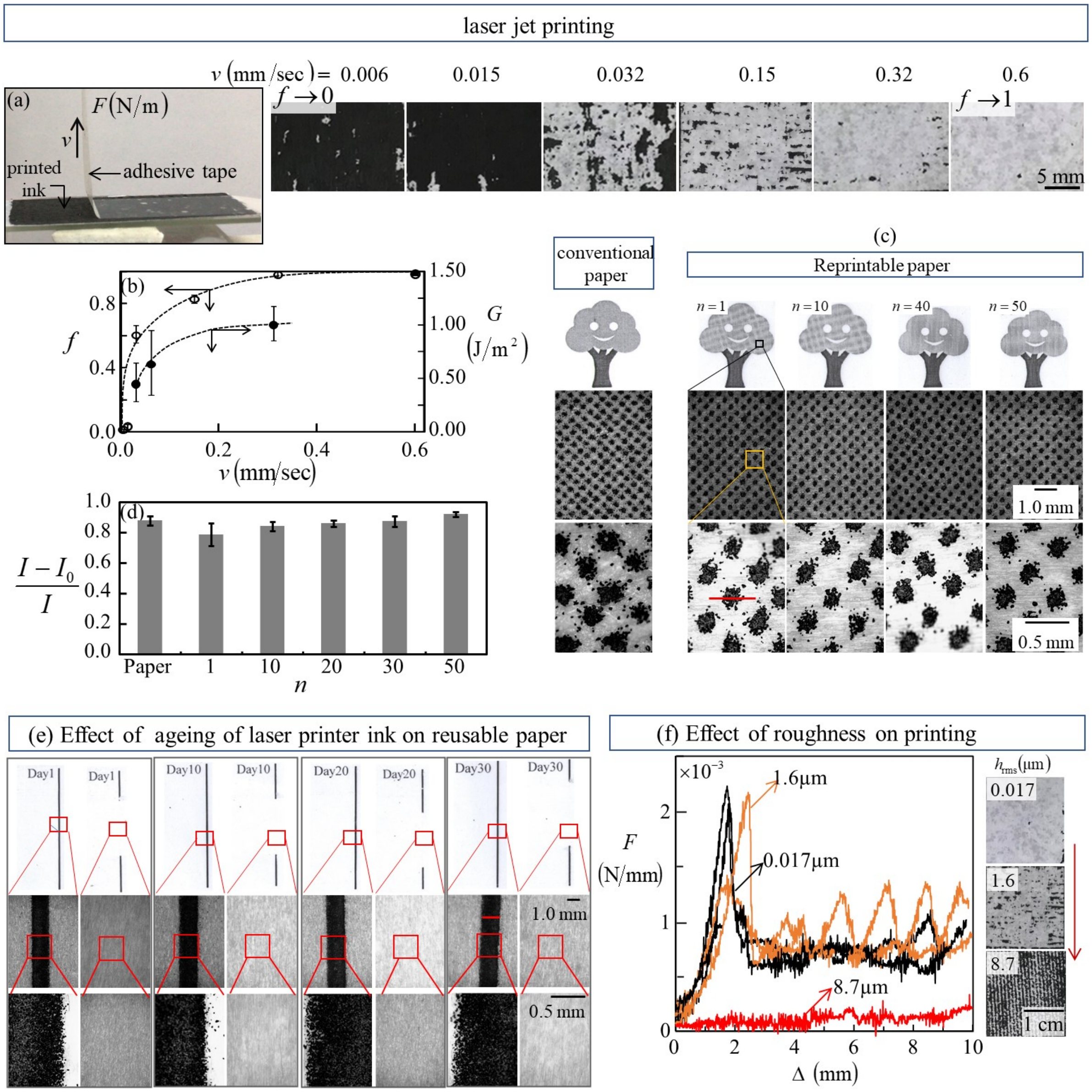

**Figure 3. (a) A reusable paper with $h_{rms} = 1.6$ μm was printed with a laser printer; a scotch tape was brought in complete contact with the print and was then peeled off at peel angle, $\theta_p = 90^o$ at different peeling speed: $v = 0.006 - 0.6$ mm sec$^{-1}$. The corresponding peeling load $F$ per unit width of the tape and the fractional area, $f$ over which the particles could be removed were measured. (b) The plot represents fracture energy, $G$, estimated as, $G = F$ and $f$ against $v$. (c) An RPS ($h_{\mathrm{rms}} = 1.6$ μm, $D_f \sim 2.11$) was subjected to printing**

**and cleaning over several cycles to demonstrate its reusability. (d) The print quality was estimated by finding the intensity, $I$ in gray scale across a printed image and then obtaining the dimensionless quantity, $(I - I_0)/I_0$. (e) The ageing effect of laser printing on the RPS ($h_{\mathrm{rms}} = 1.6\,\mu$m, $D_f \sim 2.11$) was examined by first printing a line on it and then cleaning off the ink at different time intervals. (f) The RPS of three different roughness were used for laser printing: $h_{\mathrm{rms}} = 0.017\,\mu$m, $1.6\,\mu$m and $8.7\,\mu$m. The ink was removed by peeling a scotch tape off it at 0.15mm/sec. The plot shows the corresponding force vs. displacement data.**

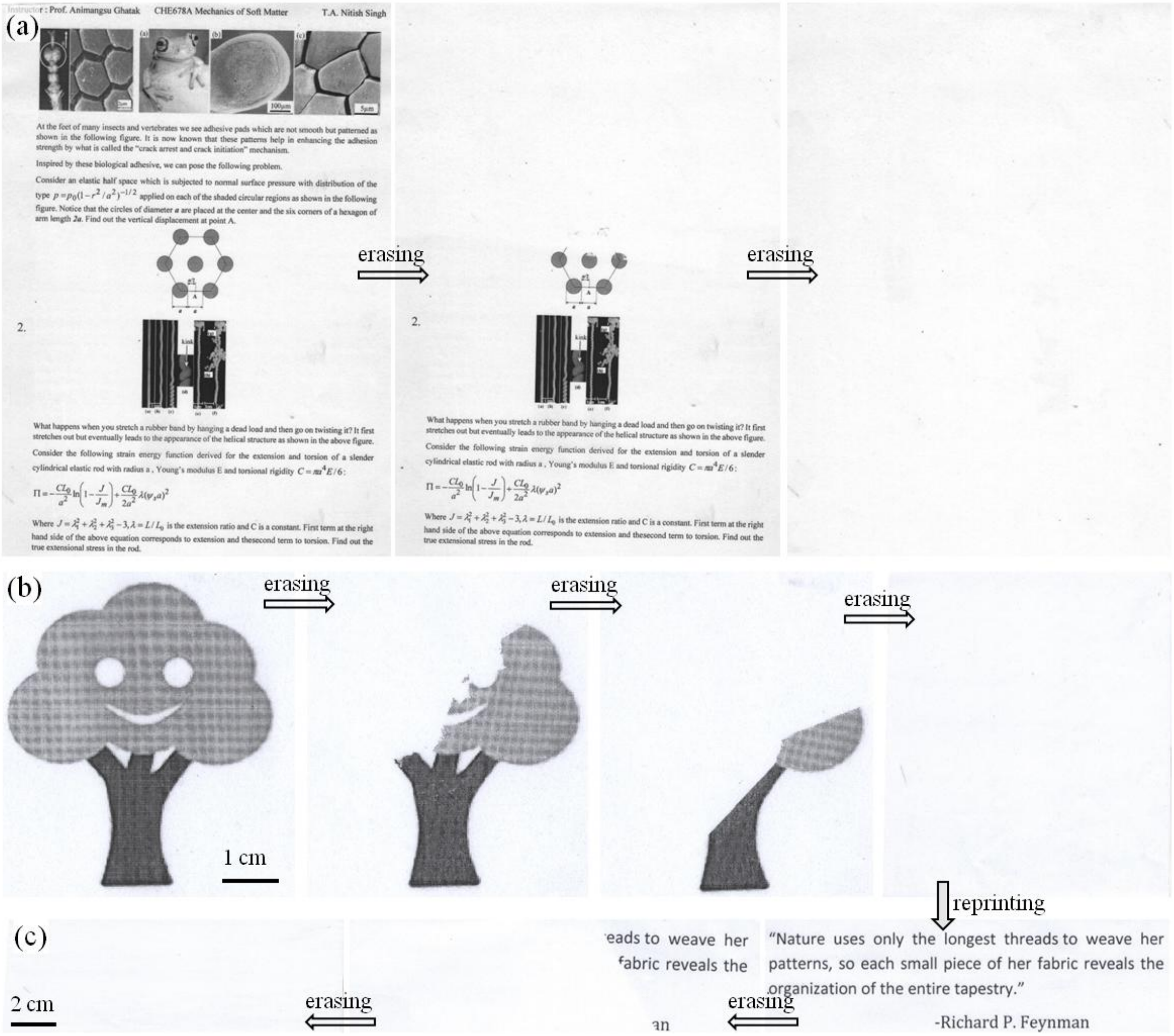

**Figure 4. The figure demonstrates reusability of RPS for printing both image and text using laser jet printer, over repeated cycles for several month. A certain document (a question paper for a course) was printed on an A4 size RPS and was stored for >2 years, during which no decline in print quality was observed. The print was removed by wiping with a cloth soaked with ethanol and the RPS was used for printing and erasing over repeated cycles.**

**Reusability of RPS for laser printing both text and image simultaneously:** To demonstrate that the RPS surfaces were good for printing both text and image over a large surface area, in figure 4(a) we present a (recently) scanned image of a typical question paper (for a course), comprising of both text and image. This paper was printed more than three years ago, and for all practical purposes, the print quality was as good as when it was freshly printed. The sequence of images shows that this paper could be completely cleaned off the ink by any of the methods described before. The images in figure 4(b) show that the same paper could then be repeatedly used for printing and erasing over repeated cycles without any deterioration of print quality.

**Printing on RPS using an inkjet printer:** In another set of experiment, the reprintability of reusable paper substrates (RPS) by an inkjet printer was probed. Figure 5(a) shows the sequence of colored images obtained over repeated cycles of printing on an RPS ($h_{\mathrm{rms}} = 1.6\,\mu\mathrm{m}$) using Epson L 1300 A3 printer and erasing. In every cycle, the image was erased by gentle wiping with a cotton cloth soaked with water. The ink could be removed also by dipping the printed paper inside a pool of water (Movie S9). Following this process, the wet RPS was dried and used again. It is worth noting that water could not form a continuous layer on the hydrophobic surface of the silicone, however a water-based ink could wet its surface and form an image. To understand this effect, a tiny drop (volume ~10μl) of black ink was dispensed on the rough surface of the silicone. Ink contains host of different chemicals including alcohols and surface-active agents, which reduce its surface energy from that of water. As a result, the advancing equilibrium contact angle of the ink drop was 52±2° (sequence of images in figure 5b) and its surface energy was estimated to be $\gamma_{ink} = 25 \pm 2\,\mathrm{mJm^{-2}}$. That of the silicone was $\gamma_{Sil} = 24.9\,\mathrm{mJ\ m^{-2}}$, so the ink could wet its surface partially. Wettability of the surface increased as the ink evaporated over time with increase in concentration of additives in it (Figure S18, Supporting Information); finally a contact angle <30° was attained which explained why the ink of inkjet printer could form a continuous film or a line over the silicone.

**Effect of surface roughness on wettability of ink:** While the above experiments were done on smooth silicone substrate, the ink droplets were dispensed also on RPS with increasing roughness. With increase in roughness of the surface, the advancing contact angle (CA) increased, albeit to a small extent (Figure 5c): from $\theta_{ink} = 53 \pm 0.5^o$ for $h_{\mathrm{rms}} = 0.017\,\mu\mathrm{m}$ it increased to $\theta_{ink} = 58.7 \pm 0.5^o$ for $h_{\mathrm{rms}} = 8.2\,\mu\mathrm{m}$. On the same set of surfaces, CA of a drop of DI water increased from $\theta_{water} = 102^o$ for smooth surface to $\theta = 120^o$ for $h_{\mathrm{rms}} = 8.2\,\mu\mathrm{m}$ corroborating with general observations that roughness increases hydrophobicity of a hydrophobic surface[48]. Experiments were done also by keeping the rough surfaces inclined to measure the dynamic advancing ($\theta_a$) and receding ($\theta_r$) CAs of ink. The data (Figure 5(d)) show that with increase in $h_{\mathrm{rms}}$ , $\theta_a$ and $\theta_r$ both varied non-monotonically while the CA hysteresis remained unaltered at ~30°±2°. While the origin of this non-monotonic behavior is not known, these observations suggest that hysteresis allowed the ink to form a continuous film in portion of surface already wetted by it. Nevertheless, the ink particles remained weakly bonded to the RPS and got easily dislodged when wiped with a damp cloth.

**Effect of dilution of ink of inkjet printer:** To understand the above effect, droplets of ink diluted with water were placed on the smooth surface of silicone. The static equilibrium contact angle (CA) of these drops show that CA increases with dilution because of increase in surface energy, $\gamma$ (Figure 5e). $\gamma$ for all different inks increased logarithmically with decrease in volume percentage of the ink in solution (Figure 5f), following the general behavior of a surface tension isotherm.

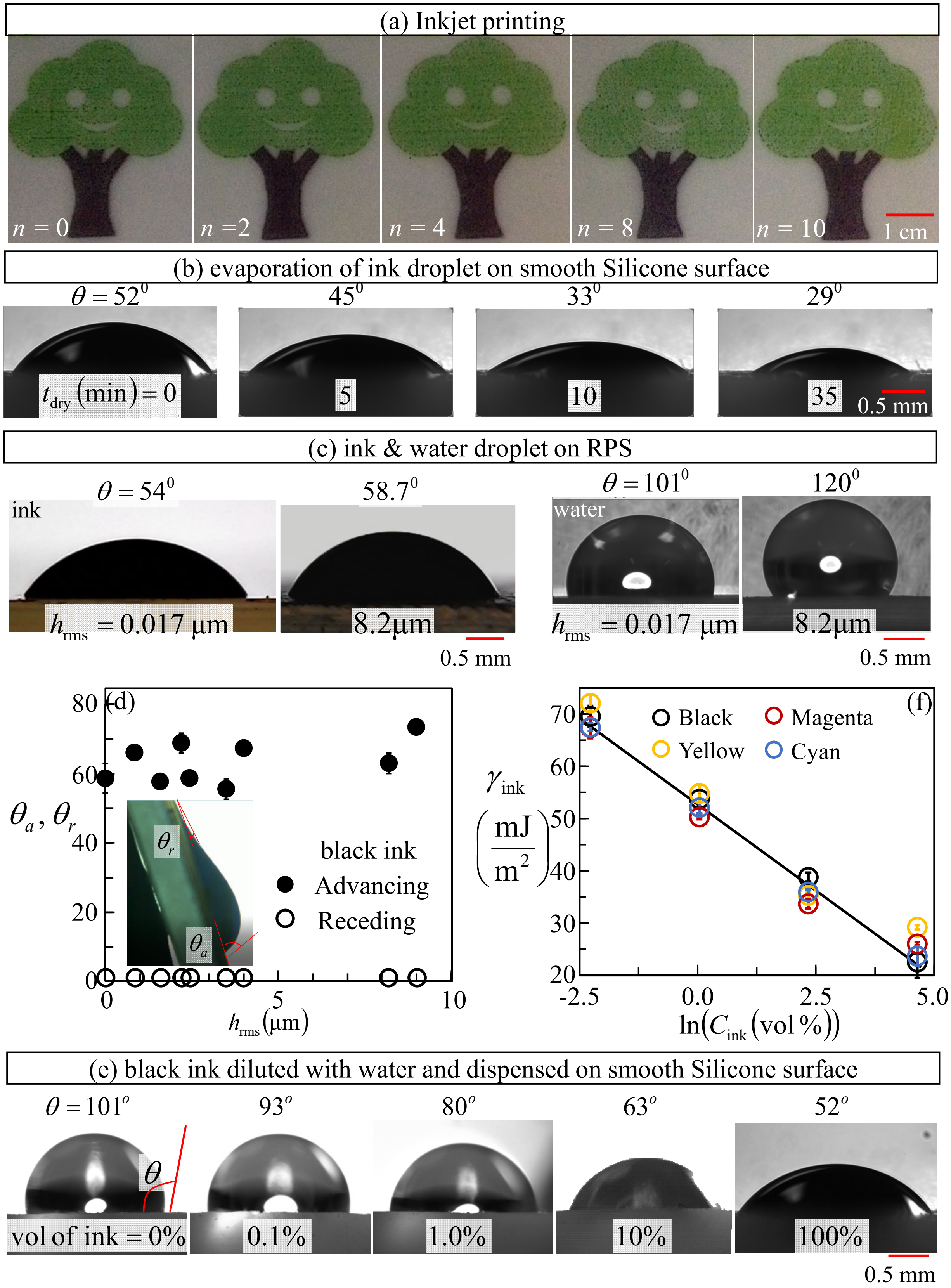
(a) Inkjet printing
n = 0
n = 2
n = 4
n = 8
n = 10
1 cm
(b) evaporation of ink droplet on smooth Silicone surface
θ = 52°
45°
33°
29°
t_dry (min) = 0
5
10
35
0.5 mm
(c) ink & water droplet on RPS
θ = 54°
58.7°
ink
h_rms = 0.017 μm
8.2μm
0.5 mm
θ = 101°
120°
water
h_rms = 0.017 μm
8.2μm
0.5 mm
(d)
θ_a, θ_r
θ_r
θ_a
black ink
Advancing
Receding
h_rms (μm)
(f)
γ_ink (mJ/m²)
Black
Magenta
Yellow
Cyan
ln(C_ink (vol %))
(e) black ink diluted with water and dispensed on smooth Silicone surface
θ = 101°
93°
80°
63°
52°
θ
vol of ink = 0%
0.1%
1.0%
10%
100%
0.5 mm

**Figure 5. The figure demonstrates use of RPS with inkjet printer. (a) A colored image was printed on an RPS ($h_{\mathrm{rms}} = 2.2\,\mu\mathrm{m}$) using the printer and the ink was cleaned off by wiping with a wet cloth. This process was repeated over several cycles. (b) A drop of black printer ink dispensed on the smooth surface of silicone formed a static advancing contact angle (CA) ~52$^{\mathrm{o}}$. Over time, because of evaporation of the solvent of the ink, CA decreased to 29$^{\mathrm{o}}$. (c) To examine the effect of roughness on the wettability of the RPS, drops of ink and DI water were dispensed on RPS of two different roughness, $h_{\mathrm{rms}} = 0.017\,\mu\mathrm{m}$ and $8.2\,\mu\mathrm{m}$. (d) To understand the effect of roughness on wetting hysteresis, droplets of the ink were dispensed on the RPS kept inclined at an angle ~70$^{\mathrm{o}}$. The advancing and receding CA ($\theta_a$ and $\theta_r$) on surfaces with different roughness, $h_{\mathrm{rms}} = 0.017 - 9.7\,\mu\mathrm{m}$, showed that hysteresis remained nearly independent of the surface roughness. (e, f) To examine the effect of dilution, a sample of black printing ink was diluted with water and drops (~10 μl) of these liquids were dispensed on the smooth surface of silicone. With increase in dilution, CA and surface energy $\gamma$ (measured by pendant drop method) of the diluted ink, both increased.**

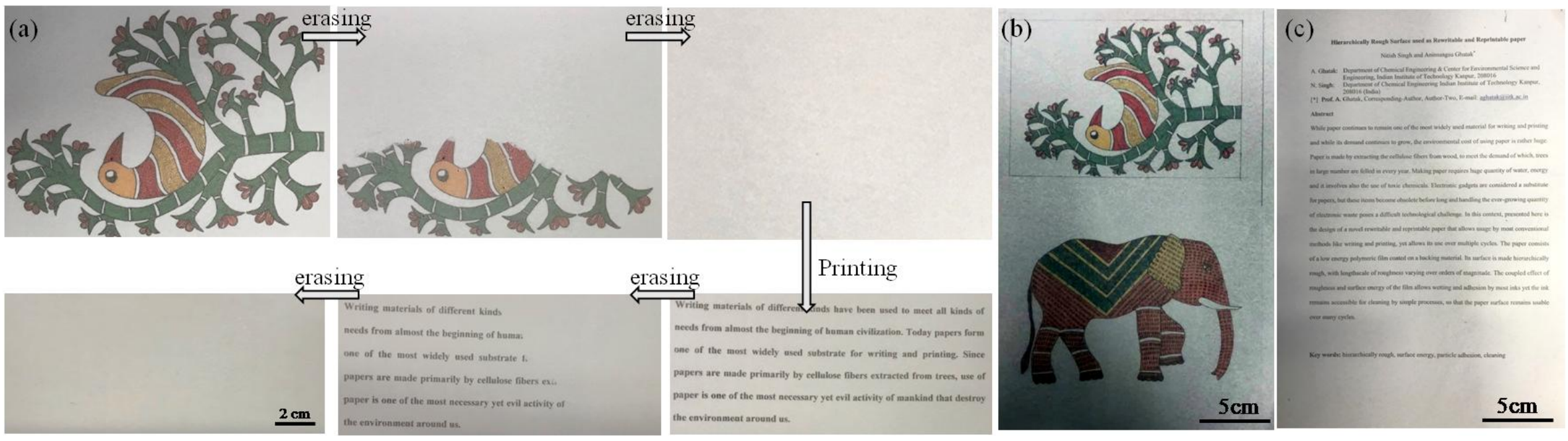

**Figure 6: Sequence of images demonstrate that the same sheet of a reusable paper can be used for printing both image and text by an inkjet printer. (a) An image was first printed with high fidelity and was then erased in steps by rinsing with water. The dried surface of**

**was used again for printing a text which too could be erased off. This sequence was repeated over several cycles as depicted in sequence (b, c).**

**Printing of both text and image using inkjet printer:** Figure 6 demonstrates that both text and images can be printed on RPS over a large area using an inkjet printer. An EPSON L 1300 SERIES printer was used to print on an A4 size RPS, for which the printer parameters were similar to that is generally used for printing on Epson premium glossy paper at high quality mode. The image sequence (and Movie S8, Supporting Information) shows also the print and erase cycle on this paper.

**Rewritability with conducting ink:** To demonstrate the generality of usage of the RPS, we have used them for writing with a conducting ink (figure 7). The ink, extracted out of a conducting ball-point pen (purchased from Circuit scribe USA) was used for drawing the outline of a typical image, with the help of a typical paint-brush. The image (of the tree) itself was drawn using conventional sketch pens. The ink was allowed to dry for 10 min. Two LED bulbs were placed at two different locations along the conducting line with the help of an adhesive. When a 9 volt battery (Godrej GP) was connected as the power source, both the bulbs glowed, indicating that a conducting line free of defect could be drawn (Figure 7a). To demonstrate that the circuit could be partially or completely removed, different parts of it were removed sequentially which caused switching off the bulbs as desired (Figure 7b,c). Finally, the whole image was cleaned off all the ink (Figure 7d); it was dried for 10 min at 40$^{o}$ C and then used again for drawing a different circuit as before (Figure 5e). The LED bulbs, positioned on the conducting line, started to glow again when the ends of the line were connected to the battery. In contrast, a line drawn on a conventional paper could not be removed by using the scotch tape (Figure S19, Movie S11, Supporting Information) or by any other method.

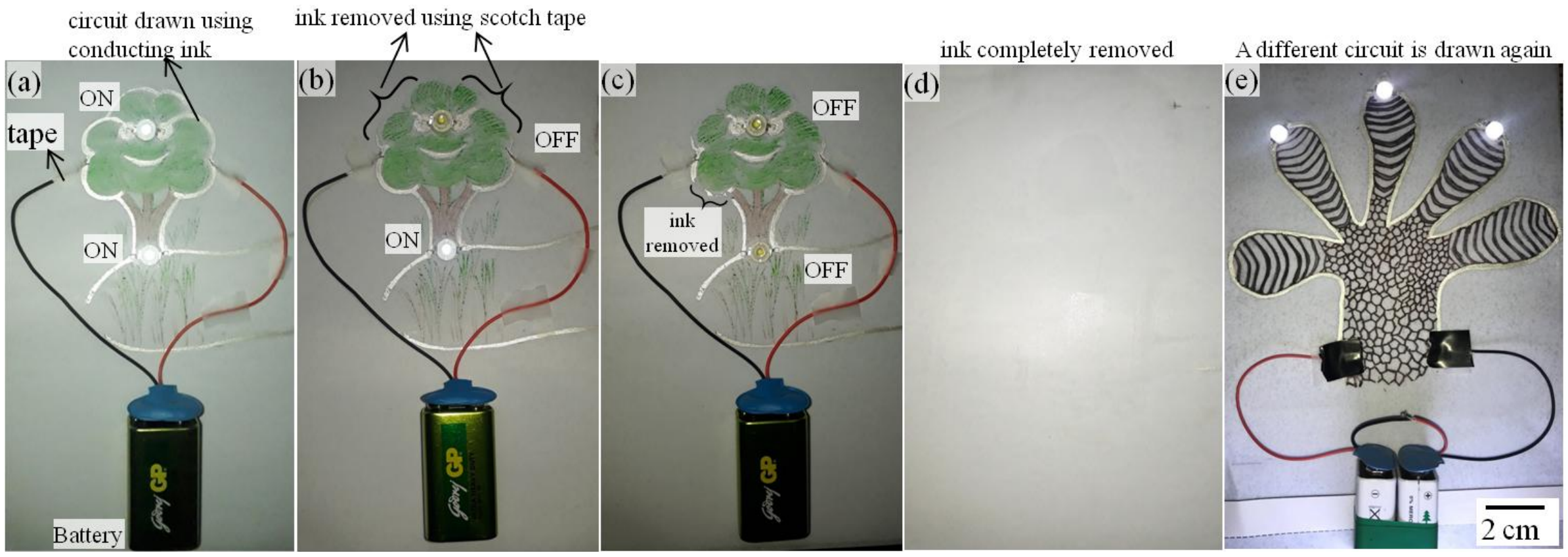

**Figure 7. The figure demonstrates reusability of RPS with a conducting ink. (a) Outline of a typical image drawn on RPS ($h_{rms} = 1.6\,\mu m$) was made using a conducting ink consisting of silver nano-flakes. Two LED bulbs, placed at two different locations on the conducting line, glowed when the two ends of the line were connected to two terminals of a battery. (b, c) Portions of the conducting line were gently peeled-off by using a scotch tape, so that the bulbs got disconnected from the battery. (d) The image was then completely removed by wiping with a damp cloth. (e) The paper was dried and reused for drawing with the conducting ink and a sketch pen ink (Luxor sign pen). A different circuit was drawn in which the LED bulbs started to glow again when the two ends of the conducting line were connected to the battery.**

**Summary:**

Recently, significant efforts have been made to fabricate “reusable” paper via synthesis of different kinds of vanishing or color-changing inks. The papers made by these technologies[49-51] remain usable over limited number of times (<20 times); the ink retention hardly exceeds 4-5 days; and none of these technologies allow both writing and printing on the same reusable paper, nor are these technologies optimized yet for large scale production. In contrast, the reusable paper substrate (RPS) presented in this report, can be used for both writing and printing via almost all conventional processes yet can be used more than 200 cycles over 2-3 years, at high fidelity and high ink retention. The unique aspect of this paper is that it can be used for printing by both laser jet and inkjet printers and can be erased as desired. The writing, printing and erasing property of the RPS can be attributed to its roughness and wettability, both of which can be varied to tune wetting and adhesion of inks of different types to the paper surface. Our one-step method for generating roughness at multiple length scales is also unique, as it combines a top-down method of casting a polymer against a template, with the bottom up approach of making the random roughness of the template itself. The resultant hierarchical roughness allows sufficient contact and adhesion with particles of size varying over wide range: 100s of nm to 10s of μm, with particles of different shape: spheroidal to ellipsoidal to arbitrarily shaped and surface roughness. Low surface energy of the RPS allows it to get wetted only partially by most aqueous inks yet its wettability increases with evaporation of the solvent. To our knowledge, this is also the first time that a conducting ink has been used to draw workable electrical circuits, reversibly and repeatedly on the reusable paper thereby opening the possibility of reusable and flexible electronic and novel paper-microfluidic applications.

**Experimental Section:**

The thin film of silicone was crosslinked against porous surface of poly(2 hydroxyethyl methacrylate) (pHEMA) as template, while a sheet of conventional paper (65-90 $gmm^{-2}$) was used as a backing for the silicone film. The pHEMA gel template was prepared by crosslinking[33] an aqueous pre-polymer solution of 2-Hydroxyethyl Methacrylate (HEMA) as the monomer, Ethylene glycol dimethacrylate (EGDMA) as cross-linker (1-2% by weight of monomer),

TEMED as a promoter (1-2% by weight of monomer), Ammonium per Sulphate (APS) as a redox initiator (0.05% by weight of monomer) and Sodium Chloride (7.8% by weight of water) as precipitant for the non-aqueous phase. The monomer-to-water weight ratio was varied such that the final product contained 80%–95% by weight of water. The gel samples once prepared were washed thoroughly with water to remove any salt and were then dried to different extent: 0-55% w/w. The silicone pre-polymer liquid mixed with the crosslinking agent was poured on this template to form a thin film. The template along with the crosslinkable liquid was placed inside a vacuum chamber to extract out any air bubble trapped inside the liquid. A sheet of conventional A4 paper was used as a backing layer. The liquid was then allowed to crosslink at room temperature, i.e. at 25° C for 4 hours and then at 60° C for 4-5 hours. The crosslinked film bonded to the backing was peeled off the template, washed with water and dried in air. The roughness of the crosslinked surface was characterized by scanning with optical profilometer (OP) (Figure S1, Supporting Information) and atomic force microscope (AFM) over different scales. The root mean square roughness, $h_{rms}$ of the silicone surface was varied from 0.39-2.2 μm (Supporting Information). For making silicone surface with roughness higher than this range ($h_{rms}$~5-12μm), sand papers of different grade were used as the template (Supporting Information). To make one with smoother surface ($h_{rms}$~0.017 μm), the silicone was crosslinked against a hydrphobized microscope glass slide.

**Conflict of Interest:** Authors declare no financial/commercial Conflict of Interest.

**Acknowledgement:** AG acknowledges Science and Engineering Research Board grant no. IMP/2018/000037 for financial assistance for this work. NS acknowledges Prateek, from Dr.

RKG's lab for help in DLS experiments and Mr. Shubham from Dr. JKS's lab for allowing us to use Goniometer for surface tension measurement.

# Hierarchically Rough Surface used as Rewritable and Reprintable paper

**Nitish Singh[1] and Animangsu Ghatak[1,2,*]**

[1] Department of Chemical Engineering, Indian Institute of Technology Kanpur, 208016
[2] Center for Environmental Science and Engineering, Indian Institute of Technology Kanpur, 208016 (India)
[*] Prof. A. Ghatak, Corresponding-Author, Author-Two, E-mail: aghatak@iitk.ac.in

## List of items

## 1. Fabrication of reusable paper substrate (RPS) reproducibility

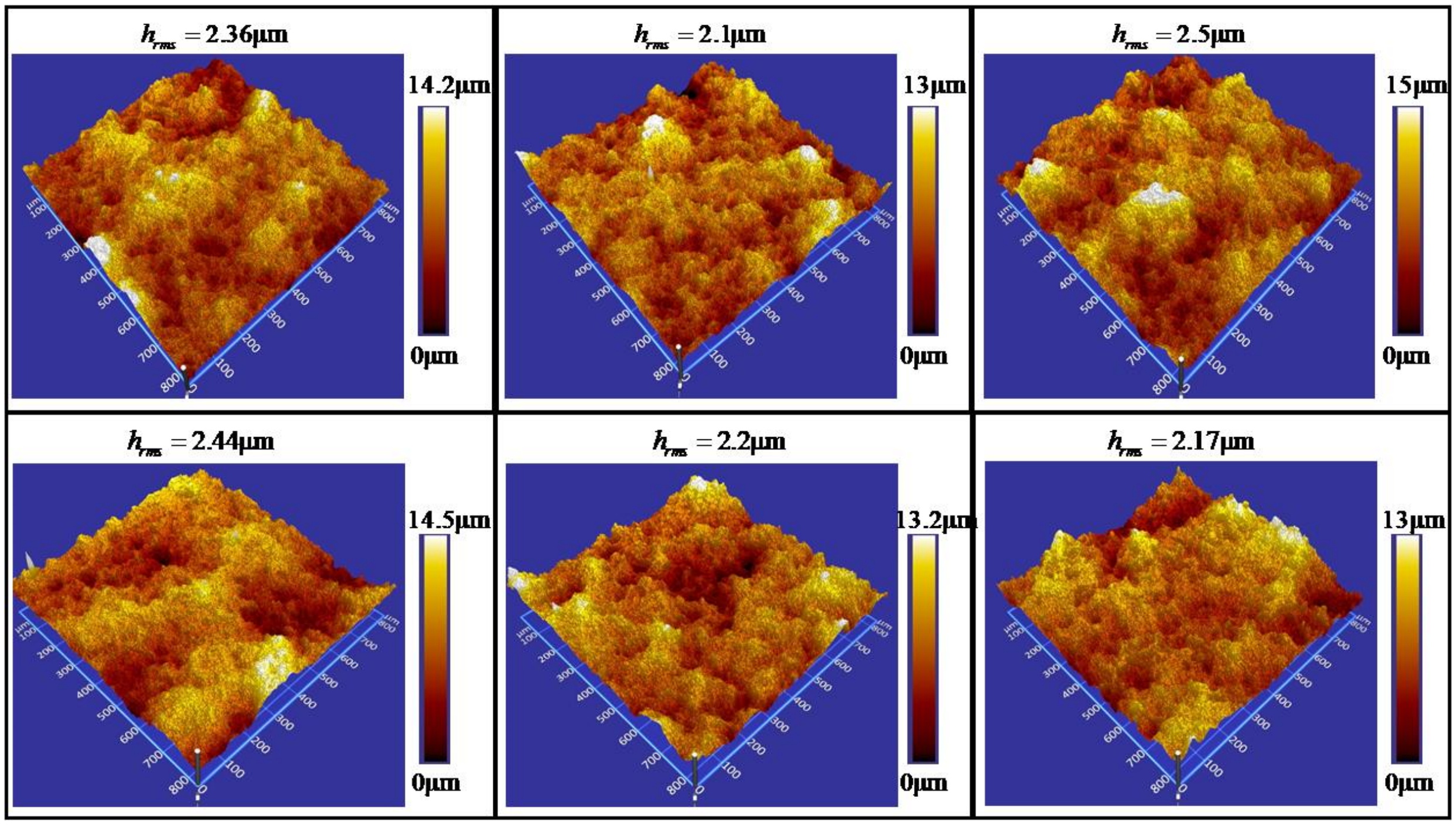

**Figure S1. To examine the reproducibility of making reusable paper substrates (RPS), several templates of the same type (pHEMA gel with 80% w/w initial water content and dried by 58%) were prepared and the silicone was crosslinked against these templates. Several RPS samples were thus made. The topography of the surface from randomly selected locations of these RPS samples was examined under optical profilometer to obtain their root mean square (RMS) roughness, $h_{\mathrm{rms}}$. The figures show examples of these optical profilometer images. The RMS roughness of these surfaces was obtained as $h_{\mathrm{rms}} = 2.36 \pm 0.23$ μm.**

**2. Estimation of surface energy of Silicone used for making the paper substrate:**

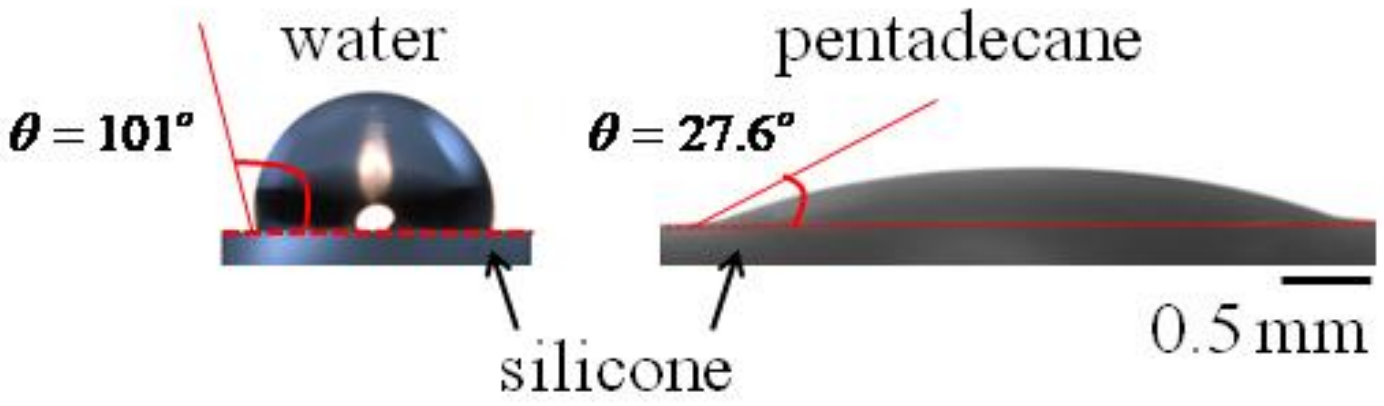

**Figure S2. Optical images depict the side view of droplets of water and pentadecane on silicone surface.**

To estimate the surface energy of the RPS surfaces, a smooth and flat surface was first prepared by crosslinking a film of the silicone against the smooth surface of a microscope glass slide. Drops of water ($\gamma_{water} = 72.8$ mJ m$^{-2}$: dispersive component, $\gamma^{d}_{water} = 21.8$ mJ m$^{-2}$ and polar component, $\gamma^{p}_{water} = 51$ mJ m$^{-2}$) and Pentadecane ($\gamma_{PD} = 27.1$ mJ m$^{-2}$: $\gamma^{d}_{PD} = 27.1$ mJ m$^{-2}$ and $\gamma^{P}_{PD} = 0$ mJ m$^{-2}$) were dispensed on it, to measure their respective contact angles: $\theta_{water-Sil} = 101^{o}$ and $\theta_{PD-Sil} = 27.6^{o}$. The Young-Dupré equation wase written for each set of solid (silicone) and liquid[1,2]:

$$(1+\cos\theta_{water-Sil})\gamma_{water} = 2(\sqrt{\gamma^{p}_{water}\gamma^{p}_{Sil}} + \sqrt{\gamma^{d}_{water}\gamma^{d}_{Sil}})$$
$$(1+\cos\theta_{PD-Sil})\gamma_{PD} = 2(\sqrt{\gamma^{p}_{PD}\gamma^{p}_{Sil}} + \sqrt{\gamma^{d}_{PDr}\gamma^{d}_{Sil}})$$

and were solved to obtain: $\gamma^{p}_{Sil} = 0.83$ mJ m$^{-2}$, $\gamma^{d}_{Sil} = 24.1$ mJ m$^{-2}$ and $\gamma_{Sil} = 24.93$ mJ m$^{-2}$.

**3. Characterization of surface roughness of Silicone surface:**

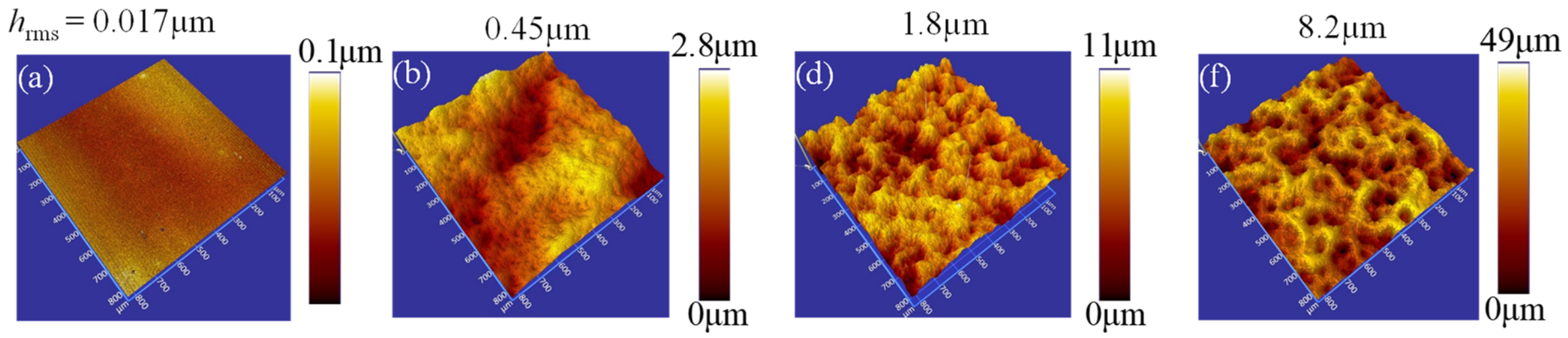

**Figure S3. (a-h) Optical profiometry images of roughness pattern on different RPS surfaces prepared using different templates.**

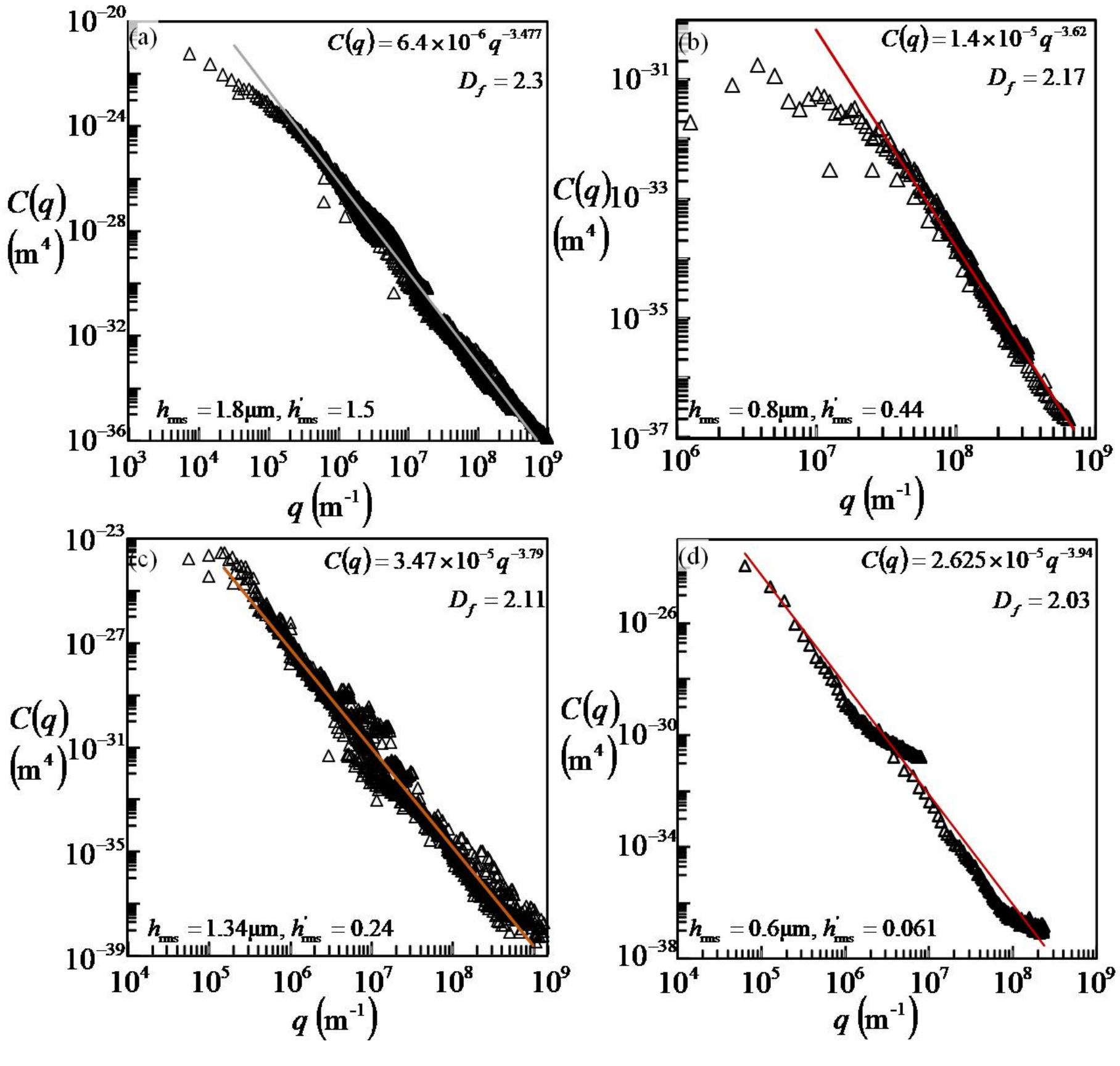

**Figure S4. Surface of RPS samples were scanned using atomic force microscopy (AFM) and optical profilometer (OP). The 2D scanned profiles obtained at different length scales: 1.7×1.7 mm$^2$, 0.8×0.8 mm$^2$ and 0.17×0.17mm$^2$ (OP) and 50×50 μm$^2$, 10×10 μm$^2$, 5×5μm$^2$,**

**$1\times1\mu m^2$, $0.5\times0.5$ $\mu m^2$ and $0.37\times0.37$ $\mu m^2$ (AFM) were used for obtaining the power spectral density (PSD) curve. For each case, 7-8 different images were analyzed and the mean value, $C(q)$ at a specific wave number, $q$ was obtained. The resultant PSD data plotted in log-log plot over different scan area, all fall on a single master line: $C(q) \sim q^{-2H-2}$, in which $H$ is known as the Hurst exponent.**

The height profile $h(x, y)$ of surfaces were used for obtaining the 2D power spectral density,

$$C(q) = \frac{1}{(2\pi)^2} \int \left\langle h(\vec{x}) h(\vec{0}) \right\rangle e^{-i\vec{q}.\vec{x}} d^2 x$$

Where, $\vec{x} = (x, y)$. To obtain the PSD for each scan area, the surface was scanned at 6-8 different locations and the PSD data for all those images were averaged to obtain the $C(q)$ vs. $q$ plot. The power spectral density (PSD) of a self-affine surface has a power law dependence on the spatial frequency, $q$ of roughness: $C(q) = c_0 q^{-2H-2}$. Here, $H$ defines the Hurst exponent and is related to the fractal dimension of the surface as, $D_f = 3 - H$. In figure S2, we show the PSD curve for different rough surfaces. As an example, the PSD data for figure S2(a) follow the relation: $C(q) \sim q^{-3.35}$. So the fractal dimension is obtained as, $D_f = 2.26$. Similarly for data in plots S2 (b-d), the fractal dimensions are found as 2.19, 2.11, 2.03. For a surface with isotropic roughness, the PSD data can be used for estimating the RMS roughness, $h_{\text{rms}}$ and the RMS slope, $h'_{\text{rms}}$ of the surface:

$$h_{\text{rms}}^2 = 2\pi \int_{q_1}^{q_2} q.C(q) dq = 2\pi \int_{q_0}^{q_{\max}} q.C(q) dq$$

$$\left(h'_{\text{rms}}\right)^2 = 2\pi \int_{q_1}^{q_2} q^3.C(q) dq = 2\pi \int_{q_0}^{q_{\max}} q^3.C(q) dq$$

Here $q_1 = q_0$ defines the roll-off wave-vector, below which the PSD curve is almost flat; $q_2 = q_{\max}$ defines the maximum resolution possible, as determined by the diameter of the AFM tip, $d_{AFM} \sim 12$ nm . Thus the RMS roughness and RMS slope are obtained as,

$$h_{\mathrm{rms}}^2 = \frac{\pi c_0}{H}\left[\frac{1}{q_0^{2H}} - \frac{1}{q_{\max}^{2H}}\right]$$

$$\left(h_{\mathrm{rms}}^{'}\right)^2 = \frac{\pi c_0}{1-H}\left[q_{\max}^{2-2H} - q_0^{2-2H}\right]$$

For $q$ varying over $10^5 < q\left(\mathrm{m}^{-1}\right) < 1.2\times10^9$, for the data plotted in figure S2(a), $h_{\mathrm{rms}}$ and $h_{\mathrm{rms}}^{'}$ are estimated as 1.8µm and 1.75 respectively. Similarly for surfaces represented by plots S2(b-d), these two quantities are estimated as (0.8µm, 0.7), (1.34 µm, 0.24) and (0.6 µm, 0.061).

**4. Preparation of gel template for generating surface with varying roughness.**

| $h_{rms}$ (μm) | Template description |
|---|---|
| 0.017 | Hydrophobized microscope glass slide |
| 0.39 | 0-2% water removal from 90%HEMA gel |
| 1.6 | 28-30% water removal from 90% HEMA gel |
| 2.2 | 55% water removal from 80% HEMA gel |
| 8.0 | 3M 400 sandpaper |

**Table S1. Surfaces with different Root mean square roughness prepared against different templates.**

| Grades of sand paper used | Surface roughness(μm) |
|---|---|
| P 2500 | 3.5 |
| P2000 | 4 |
| P1500 | 5 |
| 3M 400 | 8.2 |
| 3M320 | 9.7 |
| 3M 220 | 12 |

**Table S2. Different sand papers used for making rough surfaces.**

In another set of experiments, 3M sand-papers of grade 400,320 and 220 were used as template against which silicone was crosslinked to generate surfaces with RMS roughness 8 - 12 μm. In a similar set of experiments, sand papers of grade P2500, P2000 and P1500 (Table S2) were pressed against the surface of pHEMA (90% water content gel subjected to drying by 25-30% w/w) gel for a prolonged period (~4 hours), to transfer the roughness impression to it. This surface was then used as the template for crosslinking silicone, leading to surfaces with $h_{rms} = 3.5 - 5$ μm.

**5. Estimation of surface energy of Laser Toner particles and work of adhesion between Silicone surface and the Toner**

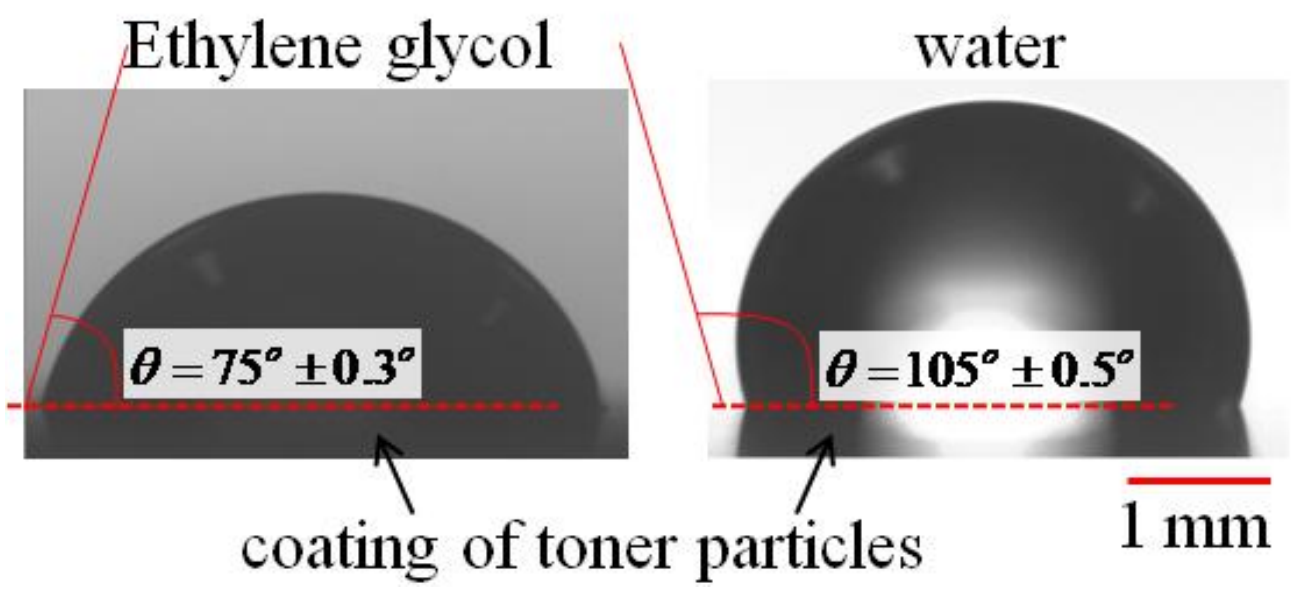

**Figure S5. Images show the side view of droplets of Ethylene glycol and water on laser printed Toner surface**.

The surface energy (SE) of Toner particles of Laser printer, $\gamma_T$ were found out to estimate their energy of adhesion, $W$ with the RPS. The ensemble average of the SE was obtained by printing a layer of the particles on a glossy paper (to minimize the effect of surface roughness). Water ( $\gamma_w = 72.8$ mJ m$^{-2}$: $\gamma_w^d = 21.8$ mJ m$^{-2}$ and $\gamma_w^p = 51.0$ mJ m$^{-2}$) and Ethylene glycol ($\gamma_{EG} = 48$ mJ m$^{-2}$: $\gamma_{EG}^d = 29$ mJ m$^{-2}$and $\gamma_{EG}^p = 19$ mJ m$^{-2}$) were used as two liquids for probing the SE. Drops of these two liquids were dispensed on this surface which formed contact angles $\theta_{EG-T} = 75^o$ and $\theta_{W-T} = 105^o$ respectively. Owens and Wendt method[1,2] was then employed, which yielded SE: $\gamma_T = 29.5$ mJ m$^{-2}$ ($\gamma_T^d = 29.4$ mJ m$^{-2}$and $\gamma_T^p = 0.1$ mJ m$^{-2}$). This value roughly matches with the values obtained earlier in literature[3].

Good-Girifalco equation[1,2] was used to estimate the interfacial energy between silicone and the Toner:

$$\gamma_{Sil-T} = \gamma_{Sil} + \gamma_T - 2(\sqrt{\gamma_T^P \gamma_{Sil}^p} + \sqrt{\gamma_T^d \gamma_{Sil}^d})$$

A theoretical estimate of the thermodynamic work of adhesion, $W = \gamma_{Sil} + \gamma_T - \gamma_{Sil-T}$ between the two adherents was thus obtained as: 54mJm$^{-2}$. It is worth pointing out that this value could not be verified experimentally, as a perfectly smooth and regularly shaped spherical toner particle as an indenter was not available. So, essentially, the value of $W$ estimated above was an ensemble average, which we round up to 60 mJ m$^{-2}$. This value was used for further calculation of actual work of adhesion using Persson's and Johnson's model of adhesion of particles on rough surfaces.

**6. Persson's model:**

For a self affine fractal surface (with Hurst exponent, $H$), by considering the excess elastic energy and the excess energy of adhesion associated with the roughness of the surface, Persson obtained the relation for actual work of adhesion as a ratio of the thermodynamic work of adhesion:

$$\frac{W_{act}}{W}=1+\frac{1}{2}(q_0h_0)^2\frac{H}{2(1-H)}\left(\left(\frac{q_{\max}}{q_0}\right)^{2(1-H)}-1\right)-\frac{E^*h_0^2q_0}{4W}\frac{H}{1-2H}\left(\left(\frac{q_{\max}}{q_0}\right)^{1-2H}-1\right)$$

Here $E^*=E/(1-\nu^2)$. It is worth noting that adhesion of particles of size $d$ is influenced by roughness patterns of wavelength $\leq d$, whereas features with wavelength $>d$ appear as smooth surface. Therefore, the minimum wave number, $q_0=2\pi/\lambda_0$ was estimated by equating $\lambda_0$ to the particle size $d$. The maximum wave number $q_{\max}$ was dictated by the diameter of the AFM tip: $q_{\max}=2\pi/d_{\mathrm{AFM}}$. Therefore, the RMS roughness $h_0$ as experienced by the particles was estimated by considering the roughness scale that varied from $d$ to $d_{\mathrm{AFM}}$.

**7. Modified Johnson's parameter for understanding particle adhesion on rough surfaces:**

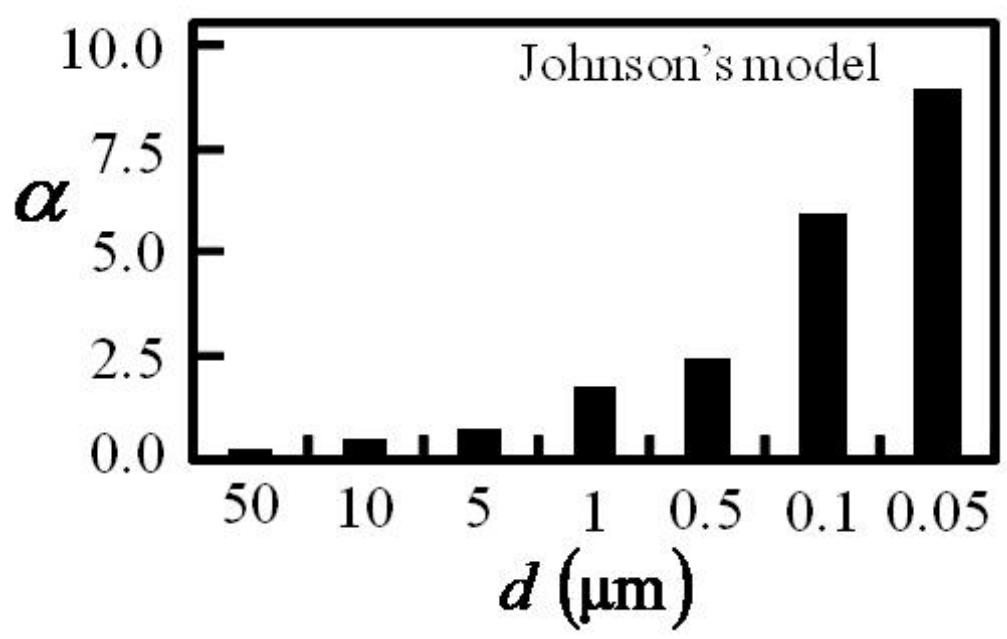

**Figure S6. Modified Johnson's parameter calculated for particles of wide ranging sizes adhering to the RPS surface as in figure 1f.**

In this approach, pioneered by Johnson[36-38], the effect of $W$ was combined with the roughness characteristics of the surface: $H$, $h_{\mathrm{rms}}$, $q_{\max}$ and $q_{\min}$ to yield a dimensionless modified Johnson's parameter:

$$\alpha = \left( \frac{4Wq_{\max}^{0.8H-1}}{\pi E^* h_{rms}^2 q_{\min}^{0.8H}} \right)^{1/2}, \text{ in which } E^* = E/\left(1-\nu^2\right)$$

For a smooth contact, $\alpha$ equates to zero; but for rougher surfaces $\alpha$ attains a finite value. Similar to Persson's model, here too, adhesion varies non-monotonically with $\alpha$. For spherical indenters, with increase in $\alpha$, adhesion initially increases weekly; in the range, $\alpha =$ 0.26-0.5, $W_{act}/W$ and correspondingly the pull-off load, $F/F_{adh}$ increases significantly (>1), beyond which it decreases with further increase in $\alpha$. In figure S5, we have estimated the $\alpha$ values for interaction of particles of different size ranges with the surface represented by the PSD curve in figure 1f ($D_f = 2.11, h_{\mathrm{rms}} = 1.08$ µm). Here, $q_{\min}$ corresponds to not the minimum wave number for the surface, but to the diameter $d$ of the particle: $q_{\min} = 2\pi/d$ .

For $W = 60$ mJ m$^{-2}$ and particles of size range $d = 50 - 5$ µm, $\alpha$ is estimated to be < 0.2, at which pull-off load is expected to be very small: $F/F_{adh} < 0.1$. For $d = 5 - 0.5$ µm, $\alpha$ is estimated to vary from 0.25 to 0.8. For such a range of values of $\alpha$, $F/F_{adh}$ has been shown to first increase and then attain a maximum value at an intermediate $\alpha$. For $d < 0.5$ µm, $\alpha$ is calculated >1, at which $F/F_{adh}$ is asymptotically decrease to 1.

**8. Method of determining particle size distribution by Dynamic light scattering (DLS) experiments:**

Particles size distributions (PSD) of ink of ball-point pen and that of inkjet printer were determined by Dynamic light scattering (DLS) using the Zetasizer Nano ZS90 (Malvern Zetasizer Nano series Nano-ZS90, Malvern Instruments Ltd.) machine. Ball-point pen ink was first diluted in ethylene glycol, followed by which it was sonicated for 20 minutes. The ink was then placed inside the cuvette, which was placed inside the cuvette holder of Zetasizer Nano ZS90. The average PSD was obtained after three runs for each sample. For determining PSD of ink of inkjet printer (Black colour from Epson), the ink was diluted with water and was subjected to the above protocol.

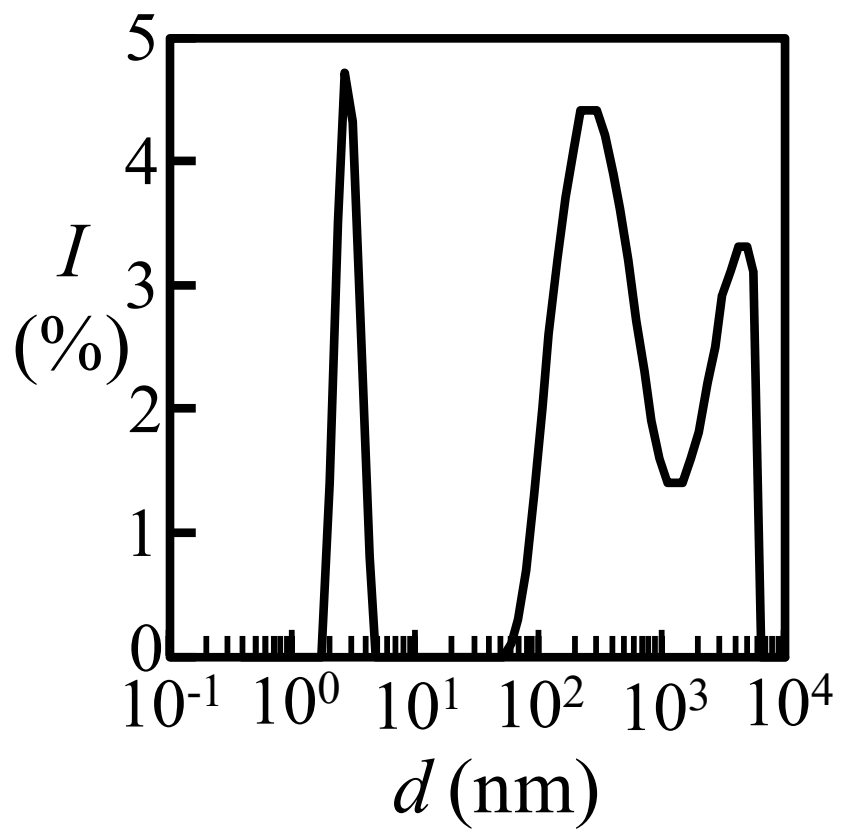

**Figure S7. PSD of black ink of inkjet printer as obtained from DLS experiments.**

**9. Estimation of work of adhesion between toner particle and silicone with an interlayer of aqueous solution of ethanol (10%w/w):**

| Surface/interface | $\gamma^{d}$ mJ m$^{-2}$ | $\gamma^{p}$ mJ m$^{-2}$ | $\gamma = \gamma^{d} + \gamma^{p}$ mJ m$^{-2}$ |
|---|---|---|---|
| Toner (1) | 29.4 | 0.1 | 29.5 |
| Silicone (2) | 24.1 | 0.83 | 24.93 |
| 10%w/w aq-Ethanol(3) | 21.03 | 14.28 | 35.21 |
| $\gamma_{12}$ (mJ m$^{-2}$) | 0.62 | | |
| $\gamma_{13}$ (mJm$^{-2}$) | 2.86 | | |
| $\gamma_{23}$ (mJm$^{-2}$) | 0.82 | | |
| $W_{132} = \gamma_{13} + \gamma_{23} - \gamma_{12}$ (mJ m$^{-2}$) | 3.06 | | |

**Table S3. Energy of different surface and interfaces.**

**10. Roughness measurement along the grid line:**

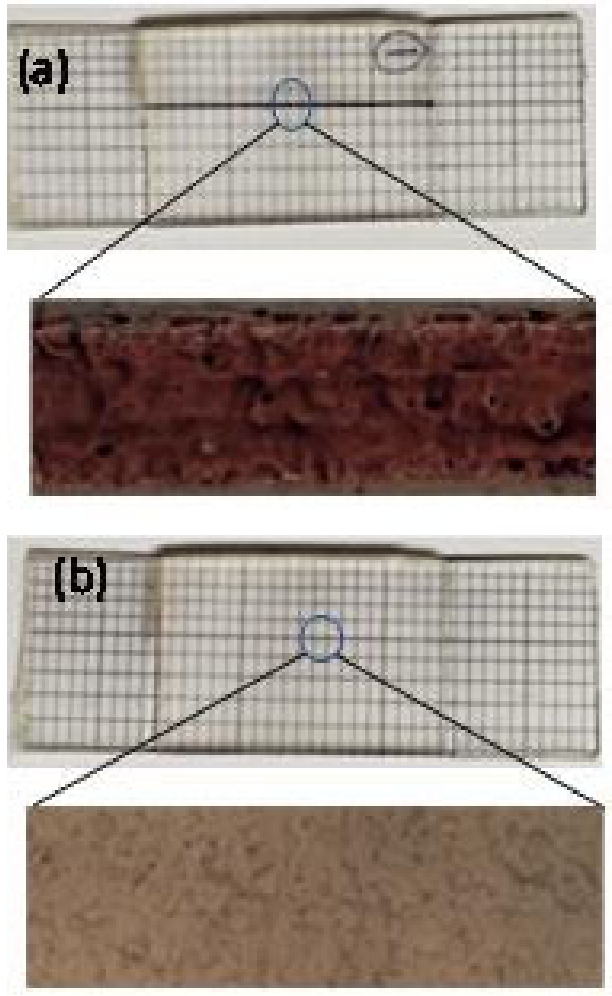

**Figure S8: (a) A typical line was drawn on the RPS using a black ball-point pen. (b) The line was erased by wiping with the wet cloth and was used again for writing. After every cycle, the roughness of the surface was measured from the optical profilometry images at five distinct points along the grid line and the average was obtained.**

**11. Smear test on line drawn on reusable paper:**

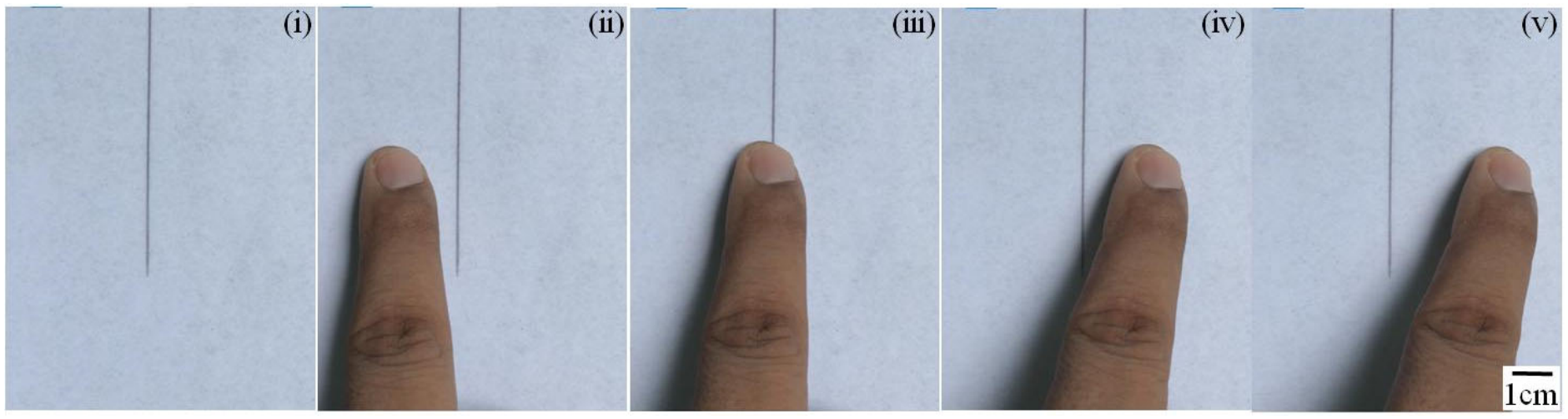

**Figure S9. A line was drawn on an RPS using a black ball-point pen (Linc Pentonic) and the ink was allowed to dry for 15 min. The RPS, at the vicinity of this line was gently sheared using finger. The dried ink was not found to get smeared off.**

**12. Optical micrograph of a line drawn on different paper substrates**

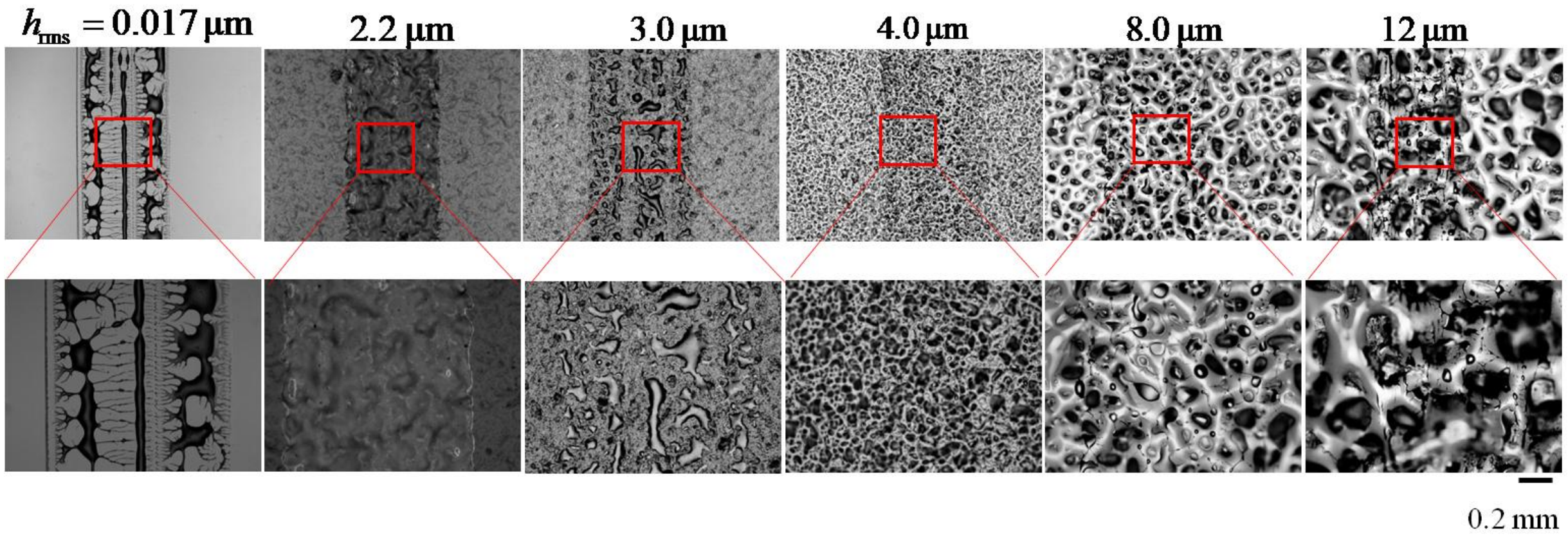

**Figure S10. Optical micrographs show the lines drawn using a black ball-point pen (GoldexKlassy Ballpoint Pen) on RPS of different RMS roughness. On RPS with small roughness: $h_{rms} = 0.017\,\mu m$, the ink layer was found to undergo instability forming a discontinuous film; on surface with $h_{rms} > 2.2\,\mu m$, the continuous line underwent Rayleigh instability and disintegrated into small patches. On surface with intermediate roughness range, $h_{rms} = 1.6$ μm, the line was found to remain continuous.**

**13. Analysis of contact line pinning following derivation by Joanny and De Gennes**

A topographical heterogeneity, like a defect, has been shown to pin a three-phase contact line of a propagating liquid front, when the slope of the surface at the vicinity of the defect exceeds a threshold limit. This limit has been shown to depend upon the surface tension, $\gamma_l$ of the liquid, intrinsic wettability of the surface measured by the contact angle (CA) $\theta_0$ of the liquid on a smooth surface of the same material and the defect size, $\delta$ with respect to the radius $r$ of the drop. For the RPS, for a line of width $w$ drawn on it, the threshold slope can be expressed as, $h'\big|_c = \left(\frac{e}{2\pi}\right)^{1/2} \frac{\pi \gamma_{ink} \theta_0^2}{\cos\theta_0 \ln(w/2\delta)}$ in which, the drop radius $r$ is replaced by $w/2$. Considering typical values of $w$ and $\delta$ as $0.5$ mm and $0.1$ mm respectively and for ink with $\theta_0 \sim 54^o \pm 1^o$ and $\gamma_{ink} \sim 39.6$ mJ m$^{-2}$ (Figure S11, Supporting Information), the threshold value of $h'\big|_c$ was estimated as ~0.21. The RMS slope $h'_{\mathrm{rms}}$, for surface with RMS roughness $h_{\mathrm{rms}} > 2\,\mu\mathrm{m}$, far exceeds this threshold limit, thereby defining a design parameter for these surfaces.

**14. Estimation of surface energy of ink of Black Ball point pen**

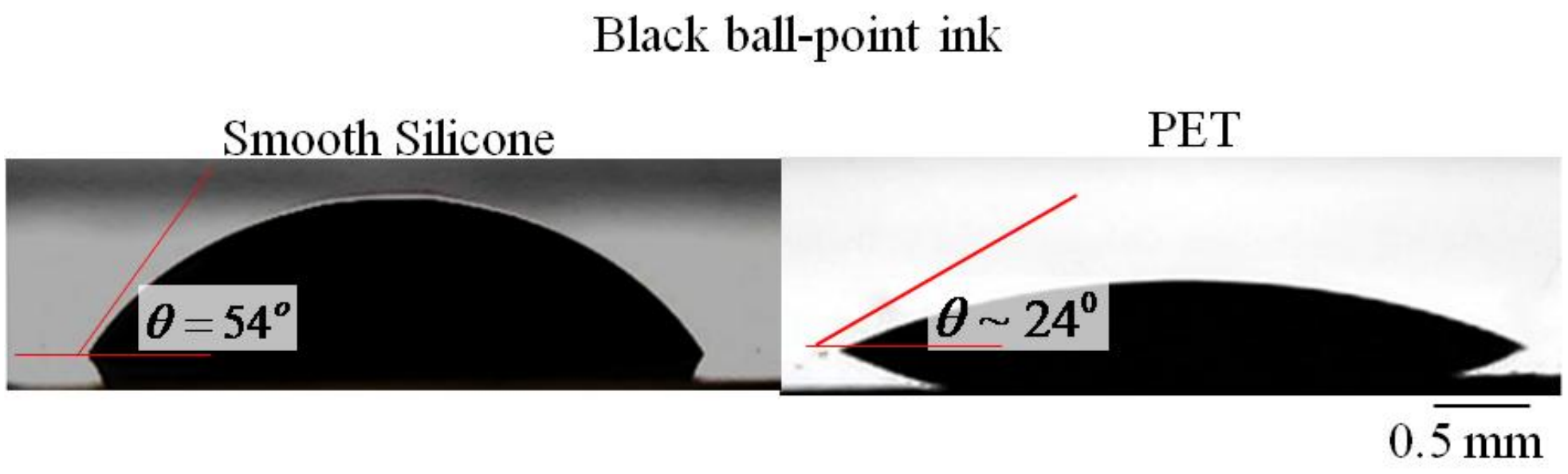

**Figure S11. Drops of ink extracted from a Reynolds ball point pen were dispensed on smooth silicone ($\gamma_{Sil} = 24.93$ mJm$^{-2}$: $\gamma_{Sil}^{d} = 24.1$ mJm$^{-2}$; $\gamma_{Sil}^{p} = 0.83$ mJm$^{-2}$) and poly(ethylene terephthalate) (PET) ($\gamma_{PT} = 36.6$ mJ m$^{-2}$: $\gamma_{PT}^{d} = 30.1$ mJ m$^{-2}$; $\gamma_{PT}^{p} = 6.5$ mJ m$^{-2}$) surfaces. The CA of the liquids and the SE values of the surfaces were used to estimate SE of the ink as $\gamma_{ink} = 39.14$ mJ m$^{-2}$.**

**15. Aging effect of adhesion of ink particles of ball-point pen on RPS**

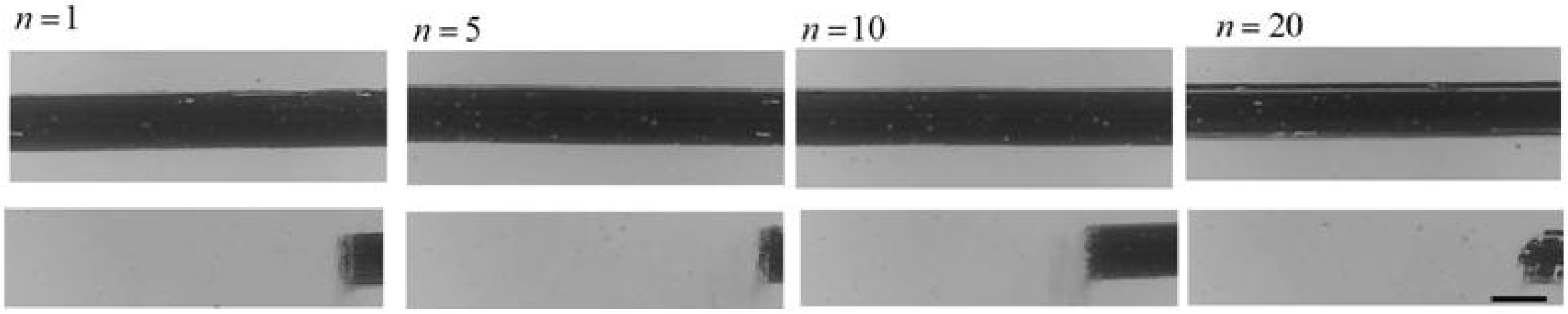

**Figure S12. A line was drawn using black Reynolds ball-point pen on an RPS with $h_{\text{rms}}=0.6$ μm and the ink was cleaned off at different time intervals. Each time a cotton cloth soaked with water was used to gently wipe off the surface. The wet surface was dried inside a hot air oven at 50° C. The erased part of the line was then observed under optical microscope. The scale bar represents 500μm.**

**16. Sketch pen used for writing on RPS**

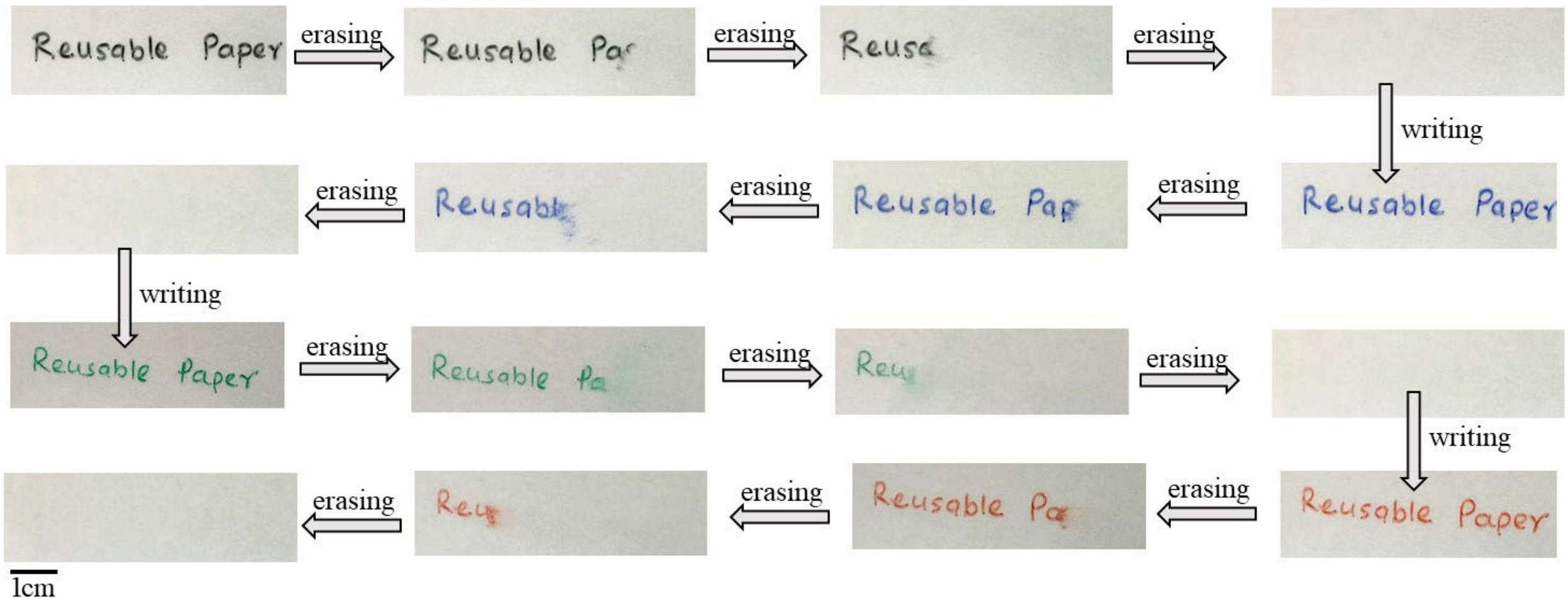

**Figure S13: Sketch pens (Luxor sign pen) of different color: Black, Blue, Green and Red were used to write on an RPS with $h_{\text{rms}} = 1.3$ μm over multiple cycles. Each time, the ink was removed by wiping with a cotton cloth damped with water and was dried at 50° C before using the RPS again for writing.**

**17. Removal of Sketch pen's ink in large scale:**

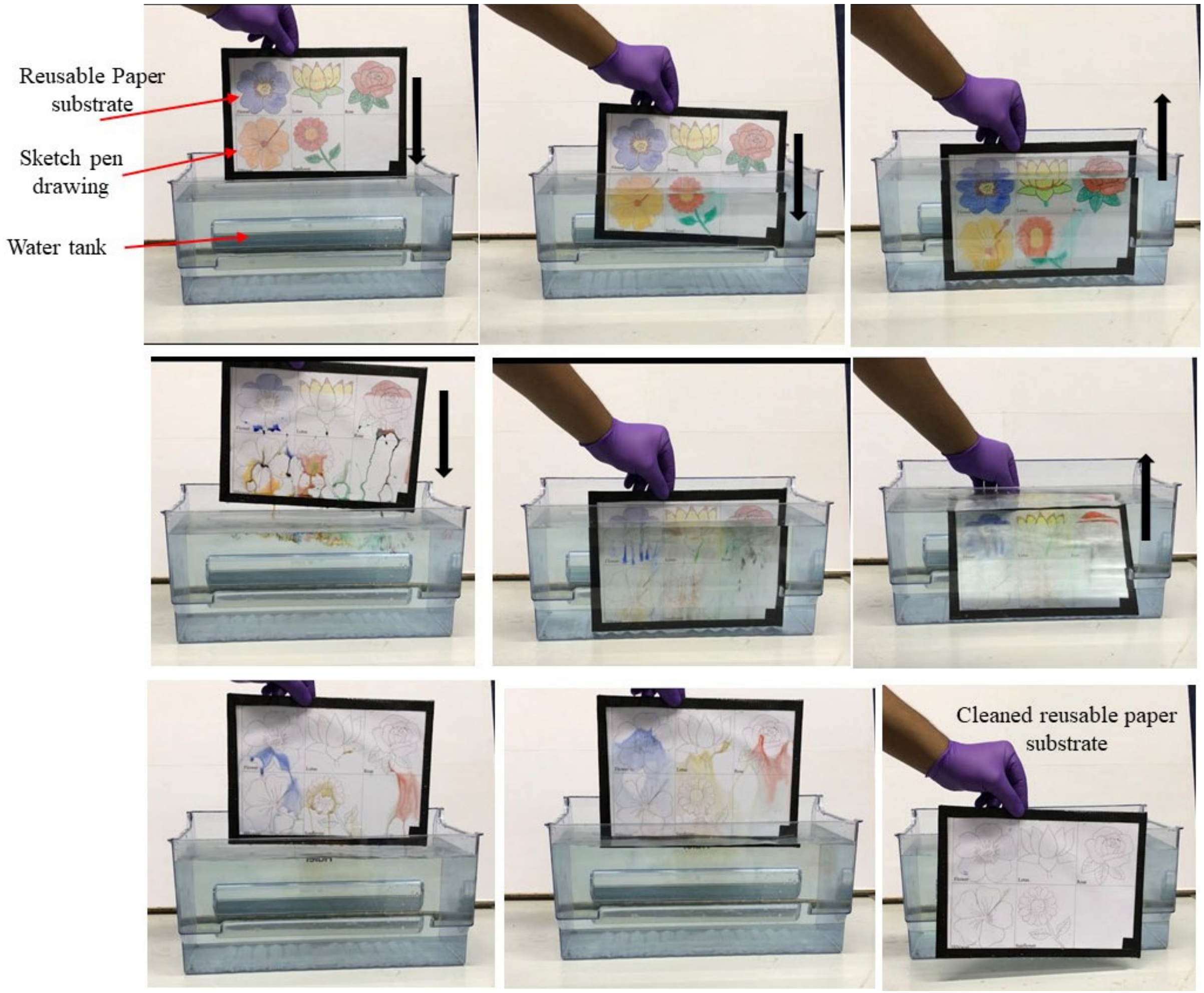

**Figure S14: Series of figures demonstrate cleaning an RPS surface of the ink of a sketch pen. The outline of a typical drawing, e.g. a flower was first printed on A4 size paper using a laser jet printer. It was then used as a backing layer of the RPS. In this process, the outline was made to remain at the background. Colours on the flowers were filled using sketch pen (Luxor sketch pen). The ink was then removed by simply submerging the sheet inside a pool of water and then withdrawing it.**

**18. Quality of RPS with respect to multiple cycle of writing and printing :**

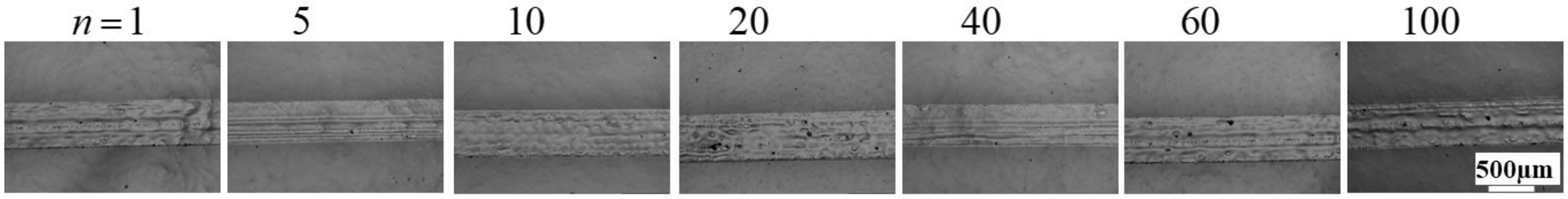

**Figure S15 (a): The sequence of images depict reusability of the RPS ($h_{\mathrm{rms}} = 0.6$ μm) over multiple cycles ($n = 1 - 100$ times) of writing with a black ball-point pen and then erasing the ink. The ink was cleaned using a damp cloth. The images show that wettability of the surface by the ink remained unaltered over multiple cycles of writing and erasing.**

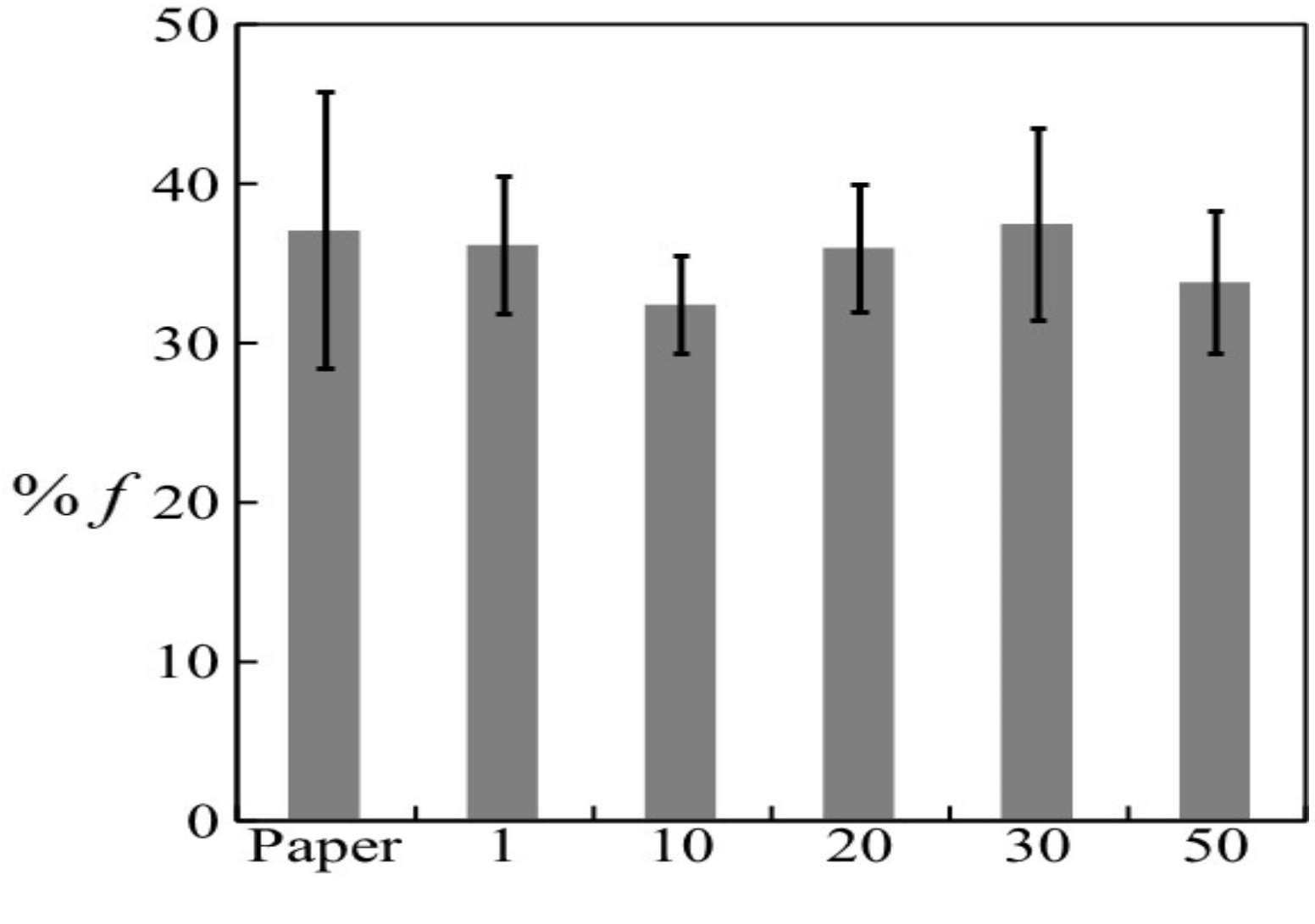

**Figure S15(b): The quality of a typical RPS ($h_{\mathrm{rms}} = 0.6$ μm) after multiple cycles of printing and erasing was determined by obtaining the area covered by toner particles after printing an image on it. A typical image as in figure 3(c) was magnified so that the dots that constitute the print could be individually identified. A circle was drawn around each such dot and the fractional area, $\%f$ covered by the toner particles within this circle was**

**estimated. Several dots were analyzed to obtain the average** $\%f$ **data. The bar chart shows that even after 50 cycles of printing and erasing,** $\%f$ **remained almost unaltered.**

**19. EDX spectra of toner particle adhered to RPS:**

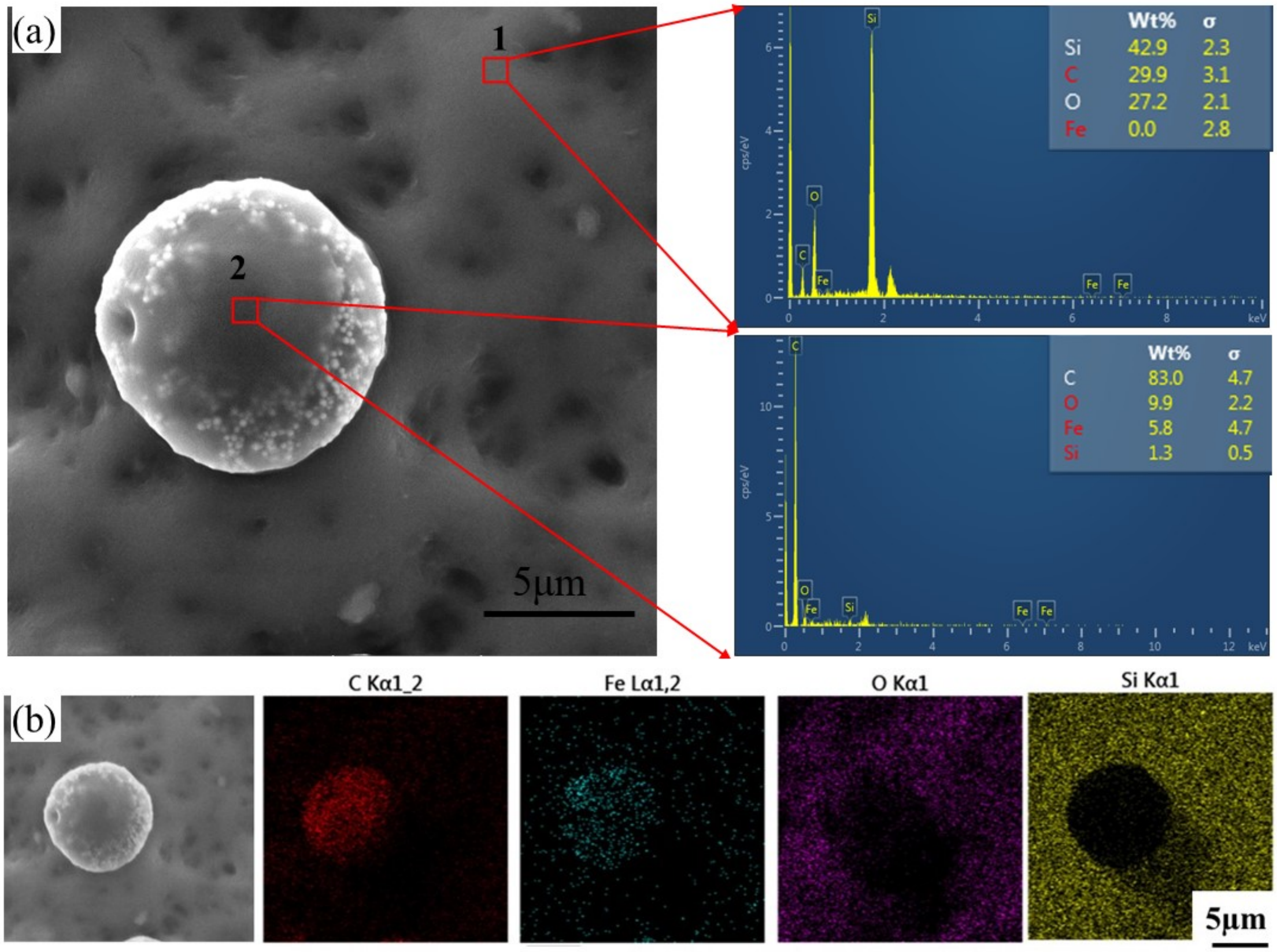

**Figure S16: (a) EDX spectra at two different locations on the RPS show that content of Fe and C at one location was significantly larger that at a different location. Similarly Si was smaller in the former than at the later location. Since, Fe and C are signature elements for toner particles, and Si is that for the silicone, these observations suggest that former location belonged to the surface of a toner particle while the later one was that of the silicone. (b) The EDX shows the typical elemental mapping of a laser jet toner particle adhered to RPS.**

## 20. SEM and EDX of before and after cleaning the RPS

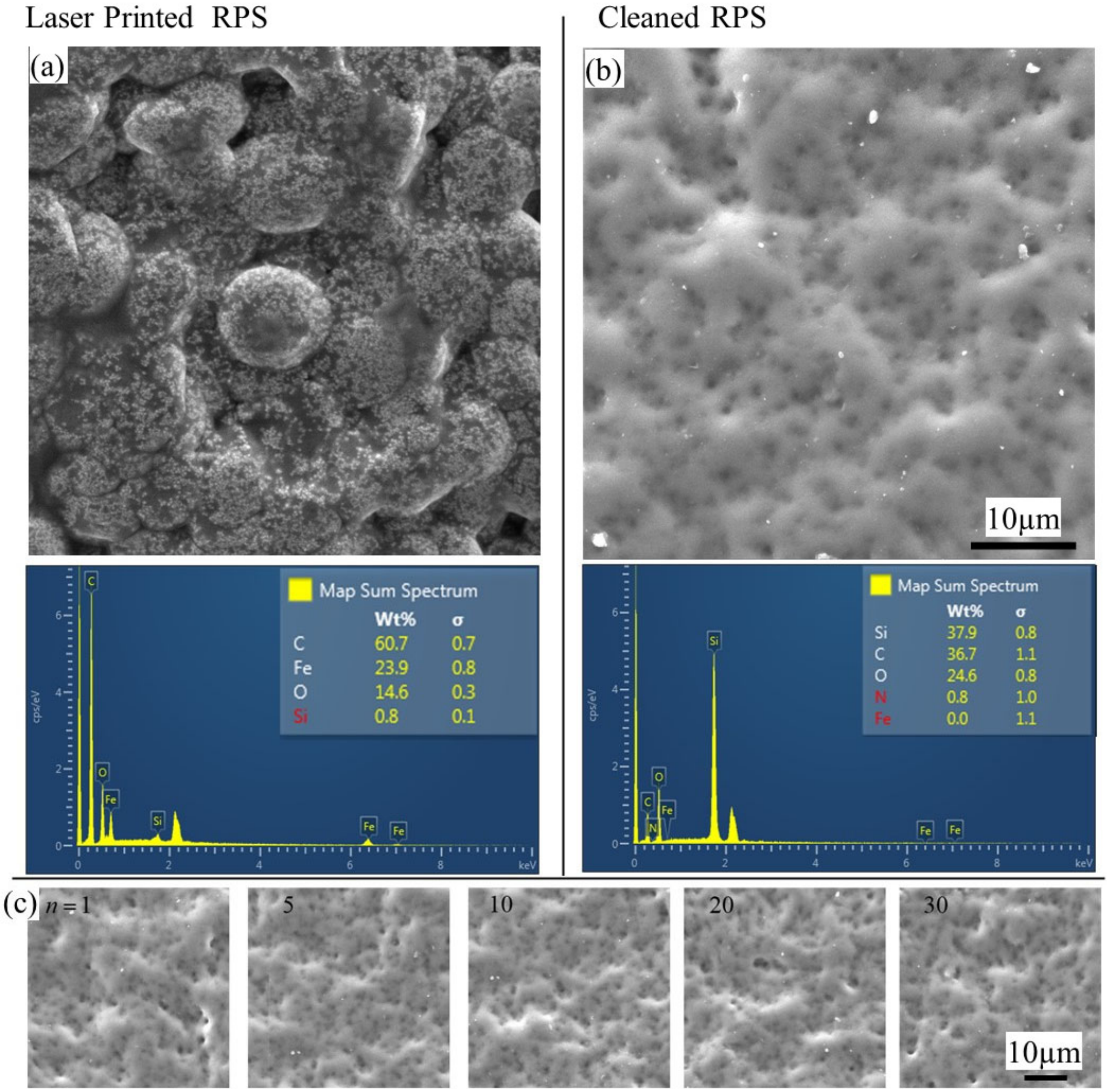

**Figure S17. (a) The figures show SEM micrograph and EDX spectra of a laser toner printed on the RPS. (b) The surface of the RPS was wiped with a cloth soaked with ethanol and was sonicated for 20-30 min in ethanol solution. Followed by it, the surface was rinsed with an aqueous solution of ethanol. The RPS was then dried at 45°C for 30 min. Dried and cleaned RPS was sputter coated with a thin layer of gold (~2-2.8nm) to make the surface**

**electrically conducting which was then taken for SEM. (c) Sequence of images shows the surface morphology of RPS ($h_{rms} = 2.23\mu\text{m}$) after multiple cycle of printing and erasing.**

**21. Retraction of ink drop during drying:**

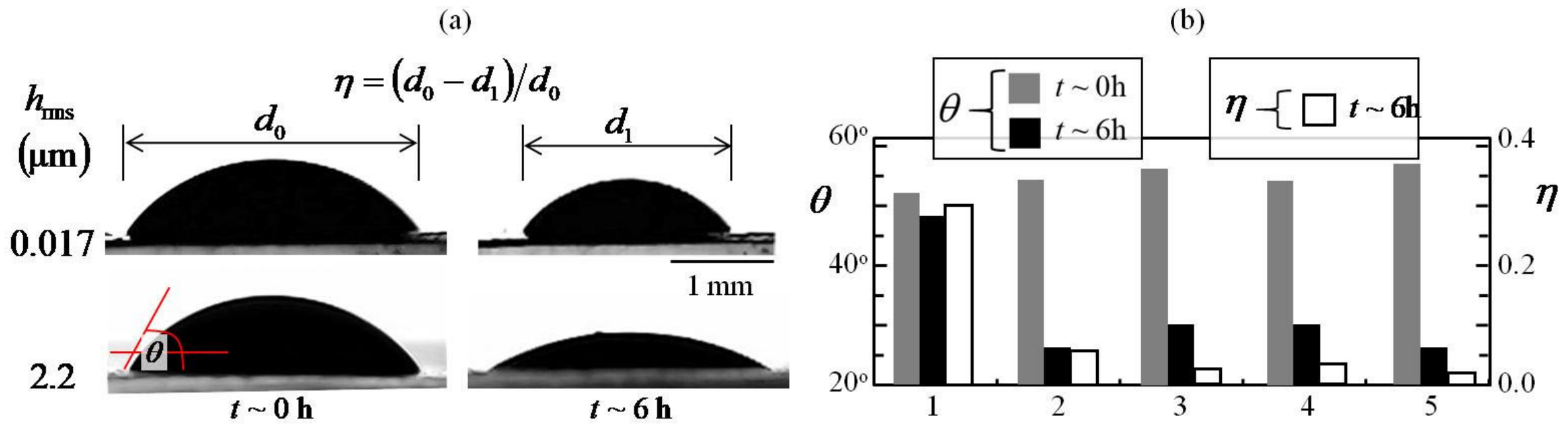

**Figure S18. (a) Drops of black ink, conventionally used in an ink-jet printer, was dispensed on RPS of two different RMS roughness, $h_{\text{rms}} = 0.017$ μm and $2.2$ μm and the drop was allowed to dry for 6 hours. The images show the initial and final configurations of the drop on these surfaces. Extent of retraction of the drop because of drying was characterized by three parameters: initial contact angle, $\theta_i$, final contact angle, $\theta_f$ and degree of retraction, $\eta$ ($\eta = \frac{d_o - d_i}{d_o}$). (b) The bar chart depicts variation of $\theta_i$, $\theta_f$ and $\eta$ with $h_{\text{rms}}$ of the RPS. Bars 1-5 represent respectively $h_{\text{rms}} = 0.017$, $1.6$, $2.2$, $5.0$ and $8.2$ μm. While the contact angles remained invariant with respect to $h_{\text{rms}}$, retraction of the three phase contact line on a smoother surface far exceeded that on the rough surface of the RPS. Smoothness led to insignificant pinning at the contact line and therefore significant retraction.**

**22. Demonstration of non-usability of a conventional 75 GSM paper with an electrically conducting ink**

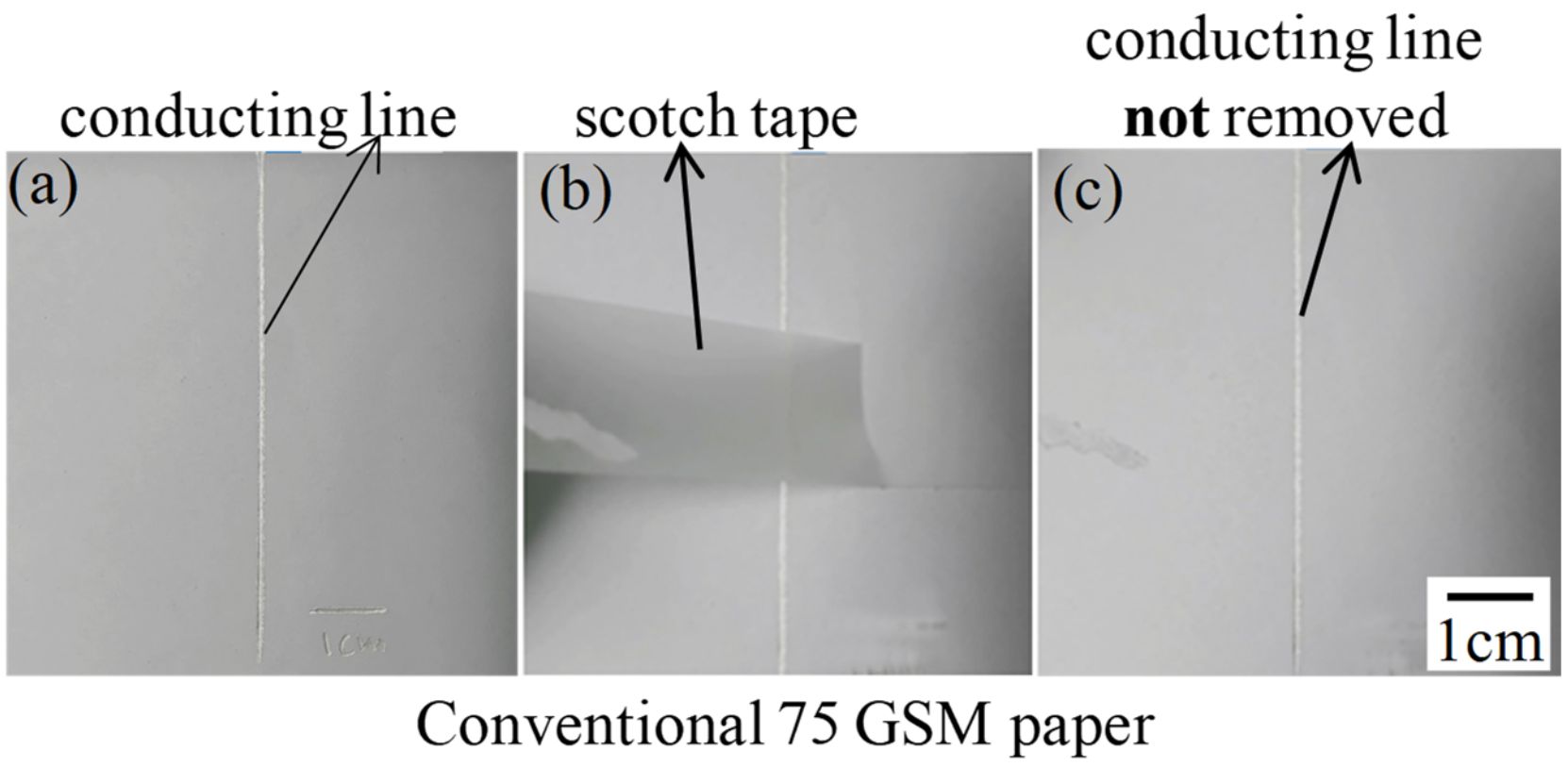

**Figure S19. The figure demonstrates non-usability of a conventional 75 GSM paper with an electrically conducting ink. Here, a line was drawn by a paint-brush using the ink available from Circuit scribe (USA) pen. The ink was then allowed to dry for ~30 min. A scotch tape was used for removing the ink by first adhering it to the surface of the paper and then peeling it off. In this process, unlike the RPS, the conventional paper got damaged because of cohesive fracture while the ink line did not get removed off the paper substrate.**

Link to movies:

https://drive.google.com/drive/folders/14rTteZp-H3a1yn5AcvSh6-pzDvPYbbdt?usp=sharing

Movie S1: The movie shows writing on an A4 size reusable paper of roughness $h_{rms} = 1.36$ µm using a black ball-point pen (Linc starline ball pen) and then removal of the ink off the paper surface by gently wiping with a piece of cotton cloth soaked with water.

Movie S2: The movie shows the write-erase cycle for writing with ball-point pens of different color: red, green, blue and black on a reusable paper of roughness, $h_{rms} = 1.36$ µm. Water soaked damp cloth was used to erase the written text after each cycle.

Movie S3: A line was drawn using a black ballpoint pen (Linc Pentonic) on a reusable paper. It was then allowed to dry for 15 min. After drying, a gentle shearing was done on the line using the index finger in order to see the stability of the line.

Movie S4: The movie shows removal of Laserjet printed toner particles from the surface of a reusable paper (roughness, $h_{rms} = 1.36$ µm) by peeling a scotch tape off it at a peeling rate of 0.32 mm/sec.

Movie S5: The movie shows the stability of Laserjet printed particles on the RPS of roughness $h_{rms} = 1.3$ µm. Stability of printed texts was checked by rubbing finger against it. The text was then removed by peeling a scotch tape off it.

Movie S6: A typical document was printed on a reusable paper of roughness, $h_{rms} = 1.6$ µm about two years ago using a Laserjet printer. The movie shows removal of toner particles from the surface of this paper by gently shearing with an Ethanol soaked damp cloth attached to a homemade indenter.

Movie S7: The movie shows printing a document on a reusable paper of roughness $h_{rms} = 1.6$ μm using a Laserjet printer and removal of toner particles off its surface using a spin scrubber (Hurricane spin scrubber) in presence of a layer of ethanol flowing over it.

Movie S8: The movie demonstrates printing and cleaning of a typical image printed using an inkjet printer on a reusable paper of roughness, $h_{rms} = 1.6$ μm. EPSON L1300 A3 color printer was used to print the colored image on the reusable paper. Printed image was then removed by rinsing water on it. A full page text was then printed on the same paper.

Movie S9: The movie demonstrates cleaning of a typical text printed using an inkjet printer on a reusable paper of roughness, $h_{rms} = 1.6$ μm. EPSON L1300 A3 color printer was used to print the text on A4 size reusable paper. The text was removed by rinsing water on it at an inclined plane.

Movie S10: The movie demonstrates the cleaning process of sketch pen's ink off an RPS. Outline of a typical drawing was first printed on an A4 size conventional paper using a Laserjet printer. The silicone coating was made on this paper with the drawing at the background to prepare the RPS. Sketch pens (Luxor sketch pen) were used to fill in colors inside the drawing. The ink was then removed by simply submerging the sheet inside a pool of water and then withdrawing from it.

Movie S11: A line was drawn on conventional A4 size paper using conducting pen (Circuit scribe, USA ). The ink was then allowed to dry for about 7 min. A transparent tape was brought in the contact with the ink on the paper and was then gently peeled off. Ink particles being adhered strongly and irreversibly on the conventional paper did not debond from it.